\documentclass[12pt]{article}
\usepackage{amsmath,cite}
\usepackage{amssymb}
\usepackage[hidelinks]{hyperref}
\usepackage{cite}
\newcommand{\p}[1]{(\ref{#1})}
\newcommand{\be}{\begin{equation}}
\newcommand{\ee}{\end{equation}}
\newcommand{\bea}{\begin{eqnarray}}
\newcommand{\eea}{\end{eqnarray}}
\newcommand{\pl}{\partial}

\newcommand{\besubeqs}{\begin{subequations}}
\newcommand{\esubeqs}{\end{subequations}}

\newcommand{\zb}{{\bar{z}}}

\newcommand{\pb}{{\bar{p}}}

\newcommand{\jb}{{\bar{j}}}

\newcommand{\pfrac}[1]{{\frac{\pl}{\pl #1}}}

\newcommand{\PP}{{\mathbb{P}}}

\newcommand{\PPb}{{\overline{\mathbb{P}}}}

\def\theequation{\arabic{section}.\arabic{equation}}
\begin{document}
\setcounter{page}0
\renewcommand{\thefootnote}{\fnsymbol{footnote}} 
\begin{titlepage}
\vskip .7in
\begin{center}
{\Large \bf  Supersymmetric Cubic Interactions For Lower Spins
From ``Higher Spin" Approach 
 } \vskip .7in 
 {\Large 
I.L. Buchbinder$^{a}$\footnote{e-mail: {\tt  joseph@tspu.edu.ru }}, 
V.A. Krykhtin$^a$\footnote{e-mail: {\tt  krykhtin@tspu.edu.ru }}, \\
M. Tsulaia$^b$\footnote{e-mail: {\tt  mirian.tsulaia@oist.jp }}, 
 D. Weissman$^b$\footnote{e-mail: {\tt  dorin.weissman@oist.jp}  }}
 \vskip .4in {$^a$ \it Center of Theoretical Physics, Tomsk State Pedagogical University, \\ 634041 Tomsk, Russia} \\
\vskip .1in { $^c$ \it Okinawa Institute of Science and Technology, \\ 1919-1 Tancha, Onna-son, Okinawa 904-0495, Japan}\\
\vskip .8in
\begin{abstract}

We demonstrate how
to reconstruct  standard cubic vertices for  $N=1$ supersymmetric Yang-Mills and 
Supergravities  using various techniques adopted for the description of cubic interactions
between higher spin fields.
\end{abstract}

\emph{\\ Based on talks given at {\it Quarks 2020} and {\it Aspects of Symmetry 2021}}

\end{center}

\vfill

\end{titlepage}

\renewcommand{\thefootnote}{\arabic{footnote}}
\setcounter{footnote}0

\section{Introduction}
The purpose of these notes is to demonstrate in
detail various approaches of constructing
supersymmetric cubic interaction vertices for
higher spin fields. To this end 
we reconstruct well known cubic vertices for
$N=1$ Super Yang-Mills and for linearized
$N=1$ Supergravities. In particular, we consider the field theoretic limits of
the  pure $N=1$, ${\cal D}=10$ supergravity \cite{Chamseddine:1980cp}--\cite{Bergshoeff:1981um},
of the $N=1$, ${\cal D}=4$ supergravity coupled with one chiral supermultiplet
\cite{Cremmer:1978hn},  
and of the $N=1$, ${\cal D}=6$ supergravity coupled
to one $(1,0)$ tensor supermultiplet \cite{Romans:1986er}--
\cite{Nishino:1986dc}.
The reason for choosing these particular types of supergravity will become clear in the following. We shall also
comment on a generalization of these vertices 
to  higher spin ``Yang-Mills-like" and ``Supergravity-
like" vertices \cite{Buchbinder:2021qrg}, and on the higher spin generalizations of the corresponding free Lagrangians \cite{Sorokin:2018djm}.

First, we shall describe the covariant BRST approach,\footnote{See 
\cite{Fotopoulos:2008ka} for a review of the BRST approach and \cite{Vasiliev:1999ba} --\cite{Tran:2020uqx} for reviews of
different approaches to higher spin theories.} which
 is similar
to the one of Open String Field Theory.\footnote{See e.g. \cite{Erler:2019vhl}
for a recent review.}
However, unlike String Field Theory, 
 the BRST approach to higher spin fields   is essentially a method
for the construction of free and interacting Lagrangians using gauge invariance as the only guiding principle,
without any recourse to a world-sheet description.

As the first step in this approach,
one constructs free Lagrangians
invariant under linear gauge transformations.
Because of the presence of gauge symmetry, these Lagrangians
contain
both physical and non-physical degrees of freedom.
Some of the non-physical degrees of freedom are removed using the equations of motion, the others are gauged away, and in the end one is left only with physical polarizations.
In general,  these systems can contain  
bosonic and fermionic fields, which are described
by Young tableaux with mixed symmetries.
However, at the free  level one can consider
 Lagrangians for just one or several (a finite number of)
  representations of the Poincar\'e group \cite{Sagnotti:2003qa}--
\cite{Francia:2002pt}.
A further requirement of supersymmetry
 singles out some particular representations of the Poincar\'e group,
both in bosonic and fermionic sectors, so the corresponding bosonic and fermionic Lagrangians are related by supersymmetry transformations.

As the second step in the BRST approach, one promotes the original gauge symmetry
to an interacting level by deforming the Lagrangian and gauge transformation
rules with nonlinear terms, in such a way that the gauge invariance is kept order by order in the coupling constant.
As in the case of free Lagrangians,  supersymmetry
singles out some particular subclass 
of the  cubic vertices,
which were  found for non-supersymmetric systems
\cite{Metsaev:2005ar}--\cite{Manvelyan:2010je}.

To describe how this approach works on the 
examples of $N=1$ Super Yang-Mills and linearized Supergravities, 
we start in Section 
\ref{LFr}
 with a description of gauge invariant free Lagrangians.
For the massless vector field the corresponding Lagrangian is the standard (Maxwell) one.
A gauge invariant Lagrangian which  contains the second rank symmetric massless tensor field
as a physical component
describes
spins $2$ and $0$ simultaneously. Then we present a similar gauge invariant description for a massless spin-vector field
 which contains irreducible representations of the Poincar\'e group with spins $\frac{3}{2}$
and $\frac{1}{2}$.  

In Section \ref{brstfree} we reformulate the results of Section \ref{LFr}
in the BRST approach. Then we impose an additional requirement of $N=1$ supersymmetry
on these systems.  Using the technique
developed in the Open Superstring Field Theory
\cite{Kazama:1986cy}
we show that $N=1$ supersymmetry requires some extra fields
(both physical and auxiliary) in the bosonic sector \cite{Sorokin:2018djm}. The obtained  Lagrangian
provides a ``unified" description
of the above mentioned $N=1$ linearized supergravities in ${\cal D}=4,6$ and $10$ dimensions.

In Section \ref{cube} we turn to cubic interactions in the covariant formalism and present generic equations
which determine the cubic vertices, as well as 
solutions to these equations
for three massless bosonic fields with arbitrary spins \cite{Fotopoulos:2010ay},
\cite{Metsaev:2012uy}.
We also present the generic equations for determining cubic vertices
for 
two massless fermions and one massless boson \cite{Buchbinder:2021qrg}.

In Section \ref{sectionsym} we describe
$N=1$ Super Yang-Mills in this formalism and
in Section \ref{sectionsugra} we consider cubic vertices for linearized supergravities. 
We comment on
 the higher spin generalization
of the vertices given in Sections \ref{sectionsym}
and \ref{sectionsugra}. 
 These cubic vertices \cite{Buchbinder:2021qrg} are covariant 
versions of the vertices for two fermionic and one bosonic
higher spin fields 
in arbitrary dimensions, 
first derived in the light cone formalism  \cite{Metsaev:2007rn}.\footnote{Covariant cubic vertices
with two fermions
and electromagnetic and gravitational fields are given in
\cite{Henneaux:2012wg}--\cite{Henneaux:2013gba}. Supersymmetric  cubic interactions on flat and $AdS$ backgrounds are given in
\cite{Alkalaev:2002rq}--\cite{Fujii:2021klx}.}

Finally, 
in   Section \ref{sectionlightcone} 
 we describe the
light cone approach to the construction of
the cubic vertices. In this approach one splits 
the generators of the Poincar\'e (super)group
into dynamical and kinematical operators.
When using the field theoretic  realization of the generators  
one takes the kinematical operators to be quadratic in terms of superfields, whereas the  dynamical generators
contain also cubic and higher order terms, i.e., cubic and higher order vertices. The requirement
that the Poincar\'e superalgebra stays intact after the nonlinear deformation
of the dynamical operators
determines these vertices order by order in the coupling constant.
We shall briefly review the construction of
\cite{Metsaev:2019dqt} (see  
 \cite{Metsaev:2019aig},\cite{Bengtsson:1983pg} for the case
of an arbitrary $N$)
for arbitrary spin supermultiplets
in  $N=1$,   ${\cal D}=1$,
and show how to obtain cubic vertices for four dimensional
$N=1$ super Yang-Mills and $N=1$
supergravity in the light cone gauge\footnote{The cubic vertices
for ${\cal D}=4$, $N=4$ super Yang-Mills in the light cone formulation are given in
\cite{Brink:1982pd}--\cite{Mandelstam:1982cb}.} as a particular example.

\section{Free Lagrangians} \setcounter{equation}0
\label{LFr}

\subsection{\texorpdfstring{$s=1$}{s=1}} \label{one}
Let us start with a massless vector field $\phi_\mu(x)$ with a standard gauge transformation rule
\be \label{lv1-1}
\delta \phi_\mu(x) = \partial_\mu \lambda(x)
\ee
The gauge invariant Klein -Gordon and transversality equations for the massless vector field can be written as 
\be \label{v1}
\Box \phi_\mu(x) = \partial_\mu C(x), \quad C(x)= 
\partial^\mu \phi_\mu(x)
\ee
where we introduced an auxiliary field  $C(x)$, which transforms 
as 
\be \label{lv1-2}
\delta C(x) = \Box \lambda(x)
\ee
The equations \p{v1} can be obtained from the Lagrangian 
\be \label{lv1}
{\cal L}= - \frac{1}{2} (\partial^\mu \phi^{\nu }) (\partial_\mu \phi_{\nu })
+ C  \partial^\mu \phi_{\mu } 
- \frac{1}{2} C^2  
\ee
After eliminating the field $C(x)$ via its own equation of motion one obtains the Maxwell Lagrangian for the vector field $\phi_\mu(x)$.

\subsection{\texorpdfstring{$s=2$}{s=2} and \texorpdfstring{$s=0$}{s=0}} \label{2and0}
One can repeat a similar consideration for a second rank symmetric tensor field
$\phi_{\mu \nu}(x)$.
Now the gauge invariant Klein-Gordon equation reads,
\be \label{s2-0}
\Box \phi_{\mu \nu}(x) = \partial_\mu C_\nu(x) + \partial_\nu C_\mu(x)
\ee
where the physical field $\phi_{\mu \nu}(x)$ and the auxiliary field $C_\mu(x)$  transform as
\be \label{s2-1-1}
\delta \phi_{\mu \nu} (x) = \partial_\mu \lambda_\nu(x) + \partial_\nu \lambda_\mu(x), \quad \delta C_\mu(x) = \Box \lambda_\mu(x)
\ee
In order to write gauge invariant transversality conditions one needs
one more auxiliary field $D(x)$, which transforms as 
\be \label{s2-1}
\delta D(x) = \partial^\mu \lambda_\mu(x)
\ee
Then the gauge invariant transversality equation is
 \be \label{s2-2}
\partial^\nu \phi_{\mu \nu}(x) - \partial_\mu D(x) = C_\mu(x) 
\ee
Finally, one can write a gauge invariant Klein-Gordon equation for the field $D(x)$
\be \label{s2-3}
\Box D(x)= \partial^\mu C_\mu(x)
\ee
and ``integrate'' the equations   \p{s2-0},  \p{s2-2} and \p{s2-3} back into the Lagrangian
 \be \label{lv2}
{\cal L}= - \frac{1}{2} (\partial^\mu \phi^{\nu \rho}) (\partial_\mu \phi_{\nu \rho})
+ 2C^\mu \partial^\nu \phi_{\mu \nu} 
-  C^\mu C_\mu  + (\partial^\mu D) (\partial_\mu D)
+ 2 D \partial^\mu C_\mu
\ee
Again, the only propagating degrees of freedom are the physical components
of the field $\phi_{\mu \nu}(x)$. Its longitudinal
components and the fields $C_\mu(x)$ and $D(x)$ are either pure gauge or zero on shell.
Finally, since there is no zero trace condition involved, one obtains a gauge invariant description
simultaneously for a spin $2$ field $g_{\mu \nu}(x)$ and for a scalar $\phi(x)$, both packed in the field $\phi_{\mu \nu}(x)$.

\subsection{\texorpdfstring{$s=\frac{3}{2}$}{s=3/2} and \texorpdfstring{$s=\frac{1}{2}$}{s=1/2}}\label{32and12}
The Lagrangian describing only spin $\frac{1}{2}$
field is simply
\be
{\cal L} = -i \, \bar \Psi \gamma^\mu \partial_\mu \Psi
\ee
The next simplest example is 
a spin-vector field $\Psi_\mu^a(x)$, where $"a"$ is a spinorial index (see appendix \ref{Appendix A} for the present conventions).
Gauge invariant Dirac and  transversality equations can be written by introducing one auxiliary field  $\chi^a(x)$ 
as
\be \label{3/2-1}
\gamma^\nu \partial_\nu \Psi_\mu (x)  +  \partial_\mu \chi (x) =0
\ee
\be \label{3/2-2}
\partial^\mu \Psi_\mu(x)  
+ \gamma^\nu \partial_\nu \chi(x) =0
\ee
 The equations \p{3/2-1}-- \p{3/2-1} are invariant under gauge transformations
 \be
\delta \Psi_\mu (x)  =  \partial_\mu \, 
 \lambda^\prime (x), 
\quad
\delta  \chi(x)  =  - \gamma^\nu \partial_\nu  \lambda^\prime (x) 
\ee
 and can be obtained from the Lagrangian
\be \label{l-f-f} 
{\cal L} = -i \, \bar \Psi^\nu \gamma^\mu \partial_\mu \Psi_\nu -
i \,   \bar \Psi^\mu \partial_\mu \chi + i \,
 \bar \chi \partial^\mu  \Psi_\mu 
+ i \, \bar \chi \gamma^\mu \partial_\mu \chi 
\ee
Similarly to the previous case, one can gauge away the auxiliary field
$\chi^a(x)$ and the non-physical polarizations of $\Psi^a_\mu(x)$. Then, one has a gauge invariant description simultaneously
of  spins  $\frac{3}{2}$ and $\frac{1}{2}$, the latter being the gamma-trace of the field 
$ \Psi_\mu^a(x)$.

\section{BRST invariant formulation} \setcounter{equation}0 \label{brstfree}
\subsection{Set Up}
A systematic way to obtain the systems described above 
is to use  the BRST approach.
Let us introduce an auxiliary Fock space spanned by one set of creation and annihilation operators. The commutation relations and the vacuum are defined in the usual way
\be
[\alpha_\mu,\alpha_\nu^+] = \eta_{\mu \nu}, \quad \alpha_\mu |0 \rangle_\alpha=0
\ee
A vector $|\Phi \rangle$ in this Fock space is a series expansion in terms of the creation operators $\alpha_\mu^+$.
In the rest of this paper
we shall take the maximal number of these oscillators 
to be equal to two. Using more than two oscillators will result in components of higher spin.
The divergence, gradient and d'Alembertian operators are realized as
\be \label{ops}
l= p \cdot  \alpha, \quad l^+= p \cdot  \alpha^{ +}, \quad l_0= p\cdot p, 
\ee
with $p_\mu = -i \partial_\mu$ and $A \cdot B \equiv \eta_{\mu\nu}A^\mu B^\nu$. The operators \p{ops} form a simple algebra with
only non-zero commutator, being
\be
[l,  l^+] = l_0
\ee
Following a standard procedure (see \cite{Fotopoulos:2008ka} for a review)
for each operator $l^+,l $ and $l_0$ one introduces ghosts $c, c^+, c_0$ of ghost number $+1$ and the corresponding
 momenta $b^+, b, b_0$ which have ghost number $-1$ . These operators obey the anticommutation relations
 \be
\{b, c^+ \} = \{b^+, c \}=\{b_0, c_0 \}=1
\ee
We define the ghost vacuum as 
 \be
 c|0 \rangle_{gh.}=b|0 \rangle_{gh.}=b_0|0 \rangle_{gh.}=0,
 \ee
 the total vacuum being $| 0 \rangle = |0 \rangle_{\alpha}\otimes |0 \rangle_{gh.} $.
 Using the corresponding nilpotent BRST charge 
 \be \label{BRST-1}
 Q = c_0 l_0 + c l^++ c^+ l - c^+ c b_0, \quad Q^2=0
 \ee
 this set up allows one to build gauge invariant free Lagrangians in a compact form,
 \be \label{LF-1}
 {\cal L}= \int dc_0 \langle \Phi |Q |\Phi \rangle
 \ee
 where the gauge transformation is given by
  \be \delta |\Phi \rangle = Q |\Lambda \rangle \ee
The Grassmann integration is carried out using the standard rule of
 \be \label{grint}
 \int dc_0 \, c_0=1
 \ee
 The requirement that the
 Lagrangian \p{LF-1} has zero ghost number uniquely fixes the expansion of an arbitrary vector $|\Phi \rangle$ 
 in terms of the ghost variables.
 Noticing  that the number operator
 \be
 N= \alpha^+ \cdot \alpha + c^+b + b^+ c
 \ee
 commutes with the BRST operator \p{BRST-1}, one has the following expansion 
 for the case of the vector field 
 \be \label{v-1}
|\Phi \rangle = (\phi^\mu(x) \alpha^{+}_\mu -i C(x)c_0 b^+ ) |0 \rangle  
\ee
Since the BRST charge has the ghost number $+1$ and the vector $|\Phi \rangle$
has the ghost number zero, then due to \p{LF-1} the parameter of gauge transformations
has the ghost number $-1$. Therefore
\be \label{lv-1}
|\Lambda \rangle = i b^+ \lambda(x)|0 \rangle
\ee
Using the equations \p{LF-1}, \p{BRST-1} and \p{v-1} one can recover the Lagrangian
\p{lv1} after performing the normal ordering and integrating over $c_0$.
Similarly, using \p{BRST-1},  \p{v-1} and \p{lv-1} one recovers the gauge transformation rules
\p{lv1-1} and \p{lv1-2}.

One can repeat the same procedure for the system described in the subsection 
\ref{2and0}. In particular, the expansion of the vector $|\Phi \rangle $
and of the parameter of gauge transformations $|\Lambda \rangle$ 
have the form
\be
|\Phi \rangle = (\phi^{\mu \nu}(x) \alpha^{+}_\mu \alpha^{+}_\nu - ic_0 b^+  C^\mu(x) \alpha^{+}_\mu 
+c^+ b^+ D(x)) |0 \rangle  
\ee
\be
|\Lambda \rangle = i b^+ \lambda^\mu(x) \alpha^{+}_\mu|0 \rangle
\ee
Using this expansion one obtains
 the Lagrangian \p{lv2} and  the gauge transformation
rules \p{s2-1-1} -- \p{s2-1}.

The BRST formulation 
for fermions is slightly more complicated, because of the anticommuting
nature of the Dirac operator
\be
g_0 = p \cdot \gamma
\ee
As a result, one has to introduce  a commuting ghost variable, which in turn
leads to an infinite expansion in its powers.
One can, however, partially fix the BRST gauge to truncate the expansion of the 
vector in the Fock space to a finite form
and write the  Lagrangian
\bea \label{ftr-free-f}
{ \cal L}_{F}&=&\frac{1}{{\sqrt 2}}\,\,{}_a\langle \Psi_{1}|(g_0)^a{}_b|\Psi_{1}\rangle^b +
 {}_a\langle \Psi_{2}|\tilde Q_{F}|\Psi_{1}\rangle^a + \\ \nonumber
&+& {}_a\langle \Psi_{1}|\tilde Q_{F}|\Psi_{2}\rangle^a +
\sqrt{2}\,\,{}_a\langle \Psi_{2}|c^+ c (g_0)^a{}_b|\Psi_{2}\rangle^b\,
\eea
where 
\be
 \label{TQ-R}
\tilde Q_{F}=
c^+_1 \, l_1   + c_1 \, l^+_1 
\ee
One can check, that the Lagrangian \p{ftr-free-f} is invariant under the gauge transformations
 \begin{eqnarray} \label{GTR1}
&& \delta\, |\Psi_{1}\rangle^a \ = \ \tilde Q_F |\Lambda^\prime \rangle^a \label{GT1}
\nonumber \\
&& \delta \,|\Psi_{2}\rangle^a \ = \ - \frac{1}{\sqrt{2}}(g_0)^a{}_b \, |\Lambda^\prime \rangle^b \  
\end{eqnarray}
and is equivalent to \p{l-f-f} 
with 
\be
|\Psi_{1}\rangle^a = \Psi^{\mu, a}(x) \alpha_\mu^+|0\rangle, \quad |\Psi_{2}\rangle^a = b^+ \chi^a (x)|0\rangle
\ee
The gauge transformations are obtained by taking the gauge parameter $|\Lambda^\prime \rangle$
to be of the form
\be
|\Lambda^\prime\rangle^a = i b^+ \lambda^{\prime a}(x) |0\rangle
\ee
As in the case of the bosonic fields, the dependence on $\alpha_\mu^+$ and on ghost variables
is uniquely fixed by the choice that the field $|\Psi_{1}\rangle^a$ contains a maximal spin
equal to $\frac{3}{2}$ and the requirement that the Lagrangian \p{ftr-free-f}
has zero ghost number.

\subsection{Supersymmetry. Linearized Supergravities} \label{linsugra}
Let us notice that the systems considered in
the subsections \ref{2and0} and \ref{32and12} can not
be connected by supersymmetry transformations,
because the  fields $\phi_{\mu \nu}(x)$
and $\Psi_\mu(x)$ have different numbers of physical degrees of freedom
on-shell.

In order to establish $N=1$ supersymmetry, one can take a formulation of the Open Superstring Field Theory  \cite{Kazama:1986cy} as a hint
and proceed as follows \cite{Sorokin:2018djm}.
Consider two independent sets
of $\alpha$-oscillators
\be
[\alpha_{\mu,m}, \alpha_{\nu, n}^+] = \eta_{\mu \nu} \delta_{mn}, \quad m,n=1,2
\ee
The corresponding divergence and gradient operators, as well as
the ghost variables $c^\pm_m$ and $b^\mp_m$ will get the index $"m"$
as well.
Therefore, we have the algebra
\be
[l_m,  l^+_n] = \delta_{mn} \, l_0 
\ee
 \be
\{b_m, c^+_n \} = \{b^+_m, c_n \}=\delta_{mn}, \quad \{b_0, c_0 \}=1
\ee
We take the fields in the fermionic sector to contain only the first set of oscillators.
In other words, we consider the system described in 
the Subsection \ref{32and12} without changes and in the corresponding
BRST formulation in the Section \ref{brstfree}
we assume that  all oscillators belong to the first set ($m=1$).

On the other hand, the vectors in the Fock space in the bosonic sector contain both types of oscillators.
Taking  physical component of the field $|\Phi \rangle$ to contain
one oscillator of each type,
we get the following expansions
\begin{eqnarray} \label{MS-1}
|\Phi \rangle&=&
 (\phi_{\mu, \nu}(x)\alpha^{\mu, +}_1 \alpha^{\nu, +}_2  -  A(x) c^+_1 b^+_2  -
B(x) c_2^+ b^+_1  \\ \nonumber
&+& ic_0 b^+_1 C_\mu(x) \alpha^{\mu, +}_2  + i c_0b^+_2 E_\mu(x) \alpha^{\mu, +}_1) 
 |0 \rangle.
\end{eqnarray}
and
\begin{eqnarray} \label{MS-2}
|\Lambda \rangle&=&
 (i b_2^+ \lambda_{\mu}(x)\alpha^{\mu, +}_1 
 +
 i b_1^+ \rho_{\mu}(x)\alpha^{\mu, +}_2
 -
  c_0b_1^+  b_2^+ \tau(x)) 
 |0 \rangle.
\end{eqnarray}
Using the corresponding nilpotent BRST charge
\be \label{BRSTMS}
Q = c_0 l_0 + \sum_{m=1,2}(c_m l_m^++ c^+_m l_m - c^+_m c_m b_0), \quad Q^2=0
\ee
it is straightforward to obtain the Lagrangian 
 \begin{eqnarray} \label{ms-l}
L_{B}&&=  - \phi^{\mu, \nu} \Box \phi_{\mu,\nu } + B\Box A  + A \Box  B \\ \nonumber
&& +E^\mu  \partial_\mu B  + C^{\nu}\partial^\mu \phi_{\nu,\mu } + C^\nu \partial_\nu  A
+ E^\mu\partial^\nu
\phi_{ \nu, \mu } \\ \nonumber
&&- B \partial_\mu E^\mu - \phi^{\nu, \mu } \partial_\mu C_\nu  -A \partial_\mu C^\mu - \phi^{\mu, \nu}
\partial_\mu E_\nu \\ \nonumber
&&+C^\mu C_\mu + E^\mu E_\mu\,.
\end{eqnarray}
by plugging the expressions \p{MS-1} and  \p{BRSTMS}
into \p{LF-1},
 performing the normal ordering of oscillators and integrating over $c_0$ according to \p{grint}.
Similarly, one can find, that  Lagrangian  \p{ms-l}
 is invariant under the gauge transformations
\bea \nonumber
&&\delta \phi_{\nu, \mu }(x) = \partial_\mu \lambda_\nu(x) + \partial_\nu \rho_\mu(x), \\ \nonumber
&&\delta A (x) = - \partial^\mu \rho_\mu(x) - \tau(x), \\ \nonumber
&&\delta B (x)= - \partial^\mu \lambda_\mu(x)+ \tau(x), \\
&&\delta C_\mu(x)  =- \Box  \lambda_\mu (x) + \partial_\mu \tau(x), \\ \nonumber
&&\delta E_\mu(x)=- \Box  \rho_\mu(x)-\partial_\mu \tau (x).
\eea
The Lagrangian \p{ms-l} is analogous to the one given in
the equation \p{lv2}. However, the present Lagrangian describes
a physical field $\phi_{\mu, \nu}(x)$ with no symmetry between
the indices $\mu$ and $\nu$. As a result, the Lagrangian  contains more auxiliary fields.
In particular, the fields $C_\mu(x)$ and $E_\mu(x)$ in \p{ms-l}
are analogous to the field $C_\mu(x)$  in \p{lv2}, and the fields
$A(x)$ and $B(x)$ are analogous to the field $D(x)$.
Again,   after eliminating the auxiliary fields 
after  gauge fixing
and using the equations
of motion one is left with only physical polarizations
in the field $\phi_{\mu, \nu}(x)$. This means,  that we have a description of a spin $2$ field $g_{\mu \nu}(x)$,
of an antisymmetric second rank tensor  $B_{\mu \nu}(x)$
and of a scalar $\phi(x)$, all contained in the field $\phi_{\mu, \nu}(x)$.

Finally, one can check that
the total Lagrangian
\bea \label{ltot}
L_{tot.}&&=  - \phi^{\mu, \nu} \Box \phi_{\mu,\nu } + B\Box A  + A \Box  B \\ \nonumber
&& +E^\mu  \partial_\mu B  + C^{\nu}\partial^\mu \phi_{\nu,\mu } + C^\nu \partial_\nu  A
+ E^\mu\partial^\nu
\phi_{ \nu, \mu } \\ \nonumber
&&- B \partial_\mu E^\mu - \phi^{\nu, \mu } \partial_\mu C_\nu  -A \partial_\mu C^\mu - \phi^{\mu, \nu}
\partial_\mu E_\nu \\ \nonumber
&&+C^\mu C_\mu + E^\mu E_\mu \\ \nonumber
&&-i \, \bar \Psi^\nu \gamma^\mu \partial_\mu \Psi_\nu -
i \,   \bar \Psi^\mu \partial_\mu \chi + i \,
 \bar \chi \partial^\mu  \Psi_\mu 
+ i \, \bar \chi \gamma^\mu \partial_\mu \chi 
\eea
being a sum of the Lagrangians
\p{l-f-f}   and \p{ms-l},
is invariant under the supersymmetry transformations
\cite{Sorokin:2018djm}
\be \label{susy-tr-1}
\delta  \phi_{ \nu, \mu }(x) = i\, \bar \Psi_{\mu}(x) \gamma_\nu \, \epsilon, \quad
\delta  C_{ \nu}(x)=  -i\, \partial_\mu \bar \chi(x)  \gamma^\mu \gamma_\nu\, \epsilon,  \quad
 \delta  B(x) = -i \,\bar \chi(x) \, \epsilon, 
\ee
\be \label{susy-tr-2}
\delta \Psi_\mu(x) = - \gamma^\nu\gamma^\rho\epsilon \,\partial_\nu \phi_{\rho,\mu}(x)
-  \epsilon E_\mu(x),
  \quad
\delta \chi(x)= - \gamma^\nu\epsilon\,  C_\nu(x)\,. 
\ee
Let us note that we have not encountered any restriction on the number of
space-time dimensions until now. The requirement that the algebra of supersymmetry transformations
\p{susy-tr-1}--\p{susy-tr-2} closes on shell
singles out the number of space-time dimensions to be
${\cal D}=3$, 4, 6, or 10.

Decomposing the fields into irreducible representations of the Poincar\'e group as
\be \label{dec-1}
\phi_{\mu, \nu}   = \left ( \phi_{(\mu, \nu )} - \eta_{\mu \nu}\frac{1}{{\cal D}} \phi^\rho{}_\rho
\right ) + \phi_{[\mu, \nu]} +
 \eta_{\mu \nu}\frac{1}{{\cal D}}  \phi^{\rho}{}_\rho
\equiv
h_{\mu \nu} + B_{\mu \nu} + \frac{1}{{\cal D}}\eta_{\mu \nu} 
\varphi 
\ee
and
\be \label{dec-2}
\psi_\mu^a = \Psi_\mu^a +\frac{1}{{\cal D}}(\gamma_\mu)^{ab} (\gamma^\nu)_{bc} \psi_\nu^c
\equiv \Psi_\mu^a + \frac{1}{{\cal D}}(\gamma_\mu)^{ab} \Xi_b
\ee
 one obtains the following $N=1$ supermultiplets:
\begin{itemize}
    \item In ${\cal D}=4$: a  supergravity multiplet $\left ( g_{\mu \nu}(x), \psi_\mu^a(x) \right )$ and a chiral multiplet
    $\left (\phi(x), a(x), \Xi(x) \right)$
    where $\partial_\mu a(x) = \frac{1}{3!}\epsilon_{\mu \nu \rho \sigma} \partial^\nu B^{\rho \sigma}(x)$.
    \item In ${\cal D}=6$:  a supergravity multiplet
     $\left ( g_{\mu \nu}(x), B_{\mu \nu}^+(x), \psi_\mu^a(x) \right )$
    and a $(1,0)$ tensor multiplet $\left (\phi(x), B_{\mu \nu}^-(x), \Xi(x) \right)$,
    where we decomposed $B_{\mu \nu}(x)$ into self-dual and anti self-dual parts.
    \item In ${\cal D}=10$: one irreducible supergravity multiplet.
\end{itemize}
Therefore, one can say, that the  Lagrangian \p{ltot} gives an ``uniform" description
of various linearized $N=1$ supergravities.
The case of ${\cal D}=3$ contains no massless propagating degrees of freedom 
with spin $2$, so we shall not consider it here.

Writing the supersymmetry transformations \p{susy-tr-1}--\p{susy-tr-2}
in terms of auxiliary oscillators,
\bea\label{STR1}
&&  \langle 0 | \delta \, \phi_{\mu, \nu}(x) \, \alpha^{\mu}_1 \, \alpha^{\nu}_2 = i\, \langle 0 |
{\overline \Psi}_{\mu}(x) \, \alpha_{1}^\mu  \,  (\gamma \cdot \alpha_2) \,  \epsilon 
\\ \nonumber
&&   \langle 0  |  \delta \, C_\mu(x) \, \alpha^{\mu}_2 \, b_1 = - \langle 0  | {\overline \chi(x)} \, g_0 \, (\gamma \cdot \alpha_2)\, \epsilon \, b_1\\ \nonumber
&&  \langle 0  | \delta \, B(x) \, b_1 \,  c_2 = - i \,\langle 0  | {\overline \chi(x)}\,  b_1 \, c_2  \, \epsilon 
\eea
\bea\label{STR2} \nonumber
&& \delta \, \Psi_{\mu}(x) \, \alpha^{\mu,+}_1|0\rangle=( -i \, g_0  \, (\gamma \cdot \alpha_2) \,  \epsilon \, 
\phi_{\mu, \nu}(x) \, \alpha^{\mu, +}_1 \, \alpha^{\nu, +}_2
- \epsilon \,E_\mu(x) \, \alpha^{\mu,+}_1)  |0 \rangle \\ 
&& \delta \, \chi (x) \, b^+_1 |0\rangle =
- (\gamma \cdot \alpha_2) \, \epsilon \,  b^+_1  \, C^ \mu(x) \, \alpha_{\mu, 2}^+ |0 \rangle
\eea
 one can see that supersymmetry  is "generated"
by the second set of oscillators  ($m=2$). In other words, to obtain the $N=1$ supermultiplets
one can start with the fermionic sector, which contains only the first
set ($m=1$), then apply the transformations \p{STR1}--\p{STR2}
and require the closure of SUSY algebra.\footnote{The same pattern persists for the higher spin supermultiplets:
the fermionic sector contains only the first set of the oscillators,
while the bosonic sector contains  the oscillators from the first set and at most one oscillator from the second set, see
\cite{Sorokin:2018djm} for the details.}

The description  for the $N=1$ supersymmetric vector multiplet is similar.
Taking the fields in the bosonic and the fermionic sectors as
\be
|\Phi \rangle=
 (\phi_{\mu}(x)\alpha_{ 2}^{\mu, +}   -i c_0b^+_2 E(x) )|0\rangle, \quad
 |\Psi \rangle= \Psi(x)|0\rangle,
\ee
one can check that the corresponding Lagrangian
\be \label{lv1-tot}
{\cal L}=  (\partial_\mu \phi^{\nu }) (\partial_\mu \phi_{\nu })
-2 E \partial^\mu \phi_{\mu } 
+ E^2 -   i \, \bar \Psi\gamma^\mu \partial_\mu \Psi
\ee
is invariant under the supersymmetry transformations
\bea \label{susyveclin}
\langle 0 | \delta \, \phi_{\mu}(x) \,  \alpha^{\mu}_2 & = & i\, \langle 0 |
{\overline \Psi}(x)   \,  (\gamma \cdot \alpha_2) \,  \epsilon 
\\ \nonumber
\delta \, \Psi(x) |0\rangle &= & 
( -i \, g_0  \, (\gamma \cdot \alpha_2) \,  \epsilon \, 
\phi_{ \mu}(x) \,   \alpha^{\mu, +}_2
- \epsilon \,E(x) )  |0 \rangle
\eea
After eliminating the auxiliary field $E(x)$
via its own equations of motion one obtains  the standard formulation
of $N=1$ vector supermultiplet in ${\cal D}= 4$, 6, or 10 with an on-shell supersymmetry.

\section{Cubic Interactions} \setcounter{equation}0
\label{cube}

\subsection{Three bosons}

In order to construct  cubic interactions for the fields considered in the previous sections,
\footnote{See \cite{Bengtsson:1987jt}, \cite{Buchbinder:2006eq} for the details of construction for higher spin fields  and
\cite{Neveu:1986mv} for the analogous 
construction in Open Bosonic String Field Theory.}
we take three copies of the auxiliary Fock space and corresponding operators. The oscillators now obey the commutation relations
\begin{equation}\label{B4-c}
[\alpha_{\mu,m}^{(i)}, \alpha_{\nu,n}^{(j), +} ] =  \delta^{ij}\delta_{mn}\eta_{\mu \nu}, 
\ee
\be
\{ c^{(i), +}_m, b^{(j)}_n \} = \{ c^{(i)}_m, b^{(j),+}_n \} 
= 
\{ c_{0,m}^{(i)} , b_{0,n}^{(j)} \} =
 \delta^{ij}\delta_{mn}\,,
\end{equation}
$$
 i,j=1,2,3, \quad m,n=1,2, \quad \mu,\nu=0,...,{\cal D}-1
$$
Then, we can consider the cubic Lagrangian
\bea \label{LNSINT}
{ \cal L}_{3B,\text{int}} &=& \sum_{i=1}^3 \int dc_0^{(i)} 
\langle \Phi^{(i)} |Q^{(i)}| \Phi^{(i)}
\rangle + \\ \nonumber 
&+&g \left ( \int dc_0^{(1)} dc_0^{(2)} dc_0^{(3)} \langle \Phi^{(1)}| \langle \Phi^{(2)}| \langle \Phi^{(3)}| |V \rangle + h.c.
\right )
\eea
where $g $ is a coupling constant and
\be \label{V3Bosons}
|V \rangle = V(p_\mu^{(i)}, \alpha_{\mu,m}^{(i),+}, c^{(i), +}_m, b^{(i), +}_m, b^{(i)}_{0,m} ) \,\, c_0^{(1)} c_0^{(2)} c_0^{(3)} \,\, 
| 0^{(1)} \rangle \otimes
|0^{(2)} \rangle \otimes | 0^{(3)} \rangle
\ee
where $V$ is a function of the creation operators that is restricted as follows. An obvious requirement is that $V$ must be Lorentz invariant. In order the Lagrangian to have the ghost number zero, the function $V$ must have the ghost number equal to zero, and finally, the requirement of the invariance of \p{LNSINT}  under the non-linear gauge transformations
\bea \label{GTNSINT}
\delta | \Phi^{(i)} \rangle & = & Q^{(i)} |\Lambda^{(i)} \rangle
-  \\ \nonumber 
&-& g \int dc_0^{(i+1)} dc_0^{(i+2)} 
\left ( (\langle \Phi^{(i+1)}| \langle \Lambda^{(i+2)}|
+ \langle \Phi^{(i+2)}| \langle \Lambda^{(i+1)}|) | V \rangle \right )
\eea
up to the first power in $g$, implies that the vertex $| V \rangle $ is BRST invariant:
\be \label{NSBRSTV}
(Q^{(1)} + Q^{(2)}+ Q^{(3)})  | V  \rangle=0
\ee
The same condition guarantees that the group structure of the gauge transformations is preserved up to the first order in $g$.
Using momentum conservation
\be \label{mcon}
p_\mu^{(1)} + p_\mu^{(2)} + p_\mu^{(3)}=0
\ee
and the commutation relations \p{B4-c},
one can show that  
that the following expressions are BRST invariant for any values of the spins entering the cubic vertex
\cite{Fotopoulos:2010ay}, \cite{Metsaev:2012uy}
\be \label{sbv-1-x}
{\cal K}^{(i)}_m = 
(p^{(i+1)}- p^{(i+2)}) \cdot \alpha^{(i),+}_m +
(b_0^{(i+1)}-b_0^{(i+2)}) \, c^{(i),+}_m,
\ee

\be \label{sbv-10}
{\cal O}^{(i,i)}_{mn} =  \alpha^{(i),+}_m \cdot \alpha^{(i),+}_n
+ c^{(i),+}_mb^{(i),+}_n + c^{(i),+}_nb^{(i),+}_m,
\ee

\be \label{sbv-3}
{\cal Z}_{mnp}= {\cal Q}^{(1,2)}_{mn} {\cal K}^{(3)}_p +
{\cal Q}^{(2,3)}_{np} {\cal K}^{(1)}_m +
{\cal Q}^{(3,1)}_{pm} {\cal K}^{(2)}_n,
\ee
where
\be \label{sbv-7}
{\cal Q}^{(i,i+1)}_{mn} = \alpha^{(i),+}_m \cdot \alpha^{(i+1),+}_n
+ \frac{1}{2} b^{(i),+}_m c^{(i+1),+}_n + \frac{1}{2} b^{(i+1),+}_n c^{(i),+}_m.
\ee

Before turning to a description of cubic vertices between bosonic and fermionic fields,
let us note 
that one can consider the cubic vertices between three bosonic fields obeying some off-shell constraints.
 In particular, 
for the fields considered in Subsection \ref{linsugra}
we shall impose  off-shell transversality conditions
\be \label{off-shell-1}
\partial^\mu\phi_{\mu, \nu}(x)=\partial^\nu\phi_{\mu, \nu}(x)=0
\ee
These conditions in turn restrict the parameters of gauge transformations 
\be \label{off-shell-2}
\partial^\mu\lambda_{\mu}(x)=\partial^\mu\rho_{\mu}(x)=0, \quad \Box \lambda_{\mu}(x)= \Box \rho_{\mu}(x)=0, \quad \tau(x)=0.
\ee
The constraints can be rewritten as
\be l^{(i)}_1 |\phi^{(i)}\rangle = l^{(i)}_2 |\phi^{(i)}\rangle = 0\,,\quad l_1^{(i)}|\Lambda^{(i)}\rangle = l_2^{(i)}|\Lambda^{(i)}\rangle = l_0^{(i)}|\Lambda^{(i)}\rangle = 0 \label{gaugefixb} \ee
As a result of these constraints, all auxiliary fields  and the ghost dependence disappears in 
$|\Phi^{(i)} \rangle$ 
and the Lagrangian 
\p{LNSINT}
reduces to
\be\label{L-bbb-gf}
{\cal L}_{\text{3B,int}} = \sum_{i=1,2,3}
\langle \phi^{(i)} |l_{0}^{(i)}| \phi^{(i)}
\rangle + 
g \left (  \langle \phi^{(1)} | \,\,
\langle \phi^{(2)}| \,\,
\langle \phi^{(3)}|  
| { V} \rangle
+ h.c \right )
\ee
The formulation in terms of the constrained fields considerably simplifies
the consideration of supersymmetry, as we shall see below.

\subsection{Two fermions and one boson} \label{subffb}

For  cubic interactions between two fermionic and one bosonic fields
 the procedure is similar. Again, in order to simplify the consideration one imposes an off-shell transversality constraint on the physical field,
 \be
 \partial^\mu \Psi_\mu^a(x)=0 \quad \Leftrightarrow \quad l_1 |\Psi\rangle = 0 \label{gaugefixf}
 \ee
  thus putting to zero the auxiliary field $\chi(x)$ (see subsection \ref{32and12}).
  This constraint, in turn, restricts the parameter of gauge transformations to
 \be
 \gamma^\mu \partial_\mu \lambda^\prime(x)=0 \quad \Leftrightarrow \quad g_0 |\Lambda\rangle = 0 \label{gaugefixf2} \ee
 The corresponding cubic Lagrangian which describes interactions
 between two fermionic and one bosonic fields has the form
 \bea \label{int.}
{\cal L}_{\text{2F1B,int}} 
&=&
\sum_{i=1}^2 
{}_a\langle \Psi^{(i)}|(g_0^{(i)})^a{}_b|\Psi^{(a)}\rangle^b + 
{}\langle \phi^{(3)}|
l_{0}^{(3)} | \phi^{(3)} \rangle + \\ \nonumber 
&&+ g\left (  \,\,    \langle \phi^{(3)} | \,\,
{}_a\langle \Psi^{(1)}| \,\,
{}_b\langle \Psi^{(2)}|  | {\cal V} \rangle^{ab}
+ h.c \right ).
\eea
The requirement that the Lagrangian \p{int.} is invariant
under the non-linear gauge transformations
\bea \label{BFFG-1} 
 \delta | \phi^{(3)} \rangle &= & 
\tilde Q_{B}^{(3)} |\Lambda_{B}^{(3)} \rangle + \\ \nonumber
&+& g ( {}_a\langle { \Psi}^{(1)}| \,\,
{}_b\langle  { \Lambda_{F}}^{(2)}|  
|{\cal W}_3^{1,2} \rangle^{ab}
+
{}_a\langle \Psi^{(2)}|  \,\, 
{}_b\langle { \Lambda}_F^{(1)}|  \
|{\cal W}_{3}^{2,1} \rangle^{ab} ),
\eea
\bea \label{BFFG-2-1}
 \delta | \Psi^{(1)} \rangle^a &= & 
{\tilde Q}_{F}^{(1)} |{ \Lambda_{F}}^{(1)} \rangle^a + 
\\ \nonumber
 &+&g   (
 {}_b\langle \Psi^{(2)}|\,\,
{}\langle  \Lambda_{B}^{(3)}|   | {\cal W}_{1}^{2,3} \rangle^{ab}
+
{ {}\langle \phi^{(3)}|\,\, 
{}_b \langle \Lambda}_F^{(2)}| |
{\cal W}^{3,2}_1 \rangle^{ab} 
),
\eea
\bea \label{BFFG-2-2}
 \delta | \Psi^{(2)} \rangle^a &= & 
{\tilde Q}_{F}^{(2)} |{ \Lambda_{F}}^{(2)} \rangle^a + \\ \nonumber
 &+&g    ({}^A\langle \phi^{(3)}| \,\,
{}_b^B\langle  \Lambda_{F}^{(1)}|   | {\cal W}_{2}^{3,1} \rangle^{ab}
+
{}_b\langle \Psi^{(1)} |\,\,
{}\langle 
{ \Lambda}_{B}^{(3)}| |
{\cal W}_2^{1,3} \rangle^{ab} 
),
\eea
as well as the requirement of preservation of the group structure
for the gauge transformations up to the first power
in the coupling constant $g$
imposes  conditions on the vertices $| {\cal V} \rangle^{ab}$ and $| {\cal W} \rangle^{ab}$  \cite{Buchbinder:2021qrg},
which are similar to \p{NSBRSTV}.

The vertex $|{\cal V}\rangle$ is again defined by a Lorentz invariant, ghost number zero function of the creation operators. Given $|{\cal V}\rangle$, gauge invariance of the Lagrangian \p{int.} holds provided one can find transformation vertices $|{\cal W}\rangle$ such that
\be \label{xy-1b}
 ({ g}_{0}^{(1)})^a{}_b | {\cal W}_1^{2,3} \rangle^{bc}
 -
 ({g}_{0}^{(2)})^c{}_b | {\cal W}_2^{1,3} \rangle^{ba}
+ 
{\tilde  Q}_{B}^{(3)} 
|{\cal V} \rangle^{ac} =0
\ee
\be \label{xy-2b}
 ({g}_{0}^{(1)})^a{}_b | {\cal W}_1^{3,2} \rangle^{bc}
+
l_0^{(3)}  
|{\cal W}_{3}^{1,2} \rangle^{ac}+
{\tilde Q}_{F}^{(2)} | {\cal V} \rangle^{ac}
=0
\ee
\be \label{xy-3b}
({g}_{0}^{(2)})^a{}_b | {\cal W}_2^{3,1} \rangle^{bc}  
+
l_0^{(3)}  
|{\cal W}_{3}^{2,1} \rangle^{ac}-
{\tilde Q}_{F}^{(1)} | {\cal V} \rangle^{ca}
=0
\ee  
The preservation of the group structure leads to another set of equations.  For consistency, there must be some functions $|{\cal X}_{i} \rangle$ such that
\be \label{inv-w-1b}
 {\tilde Q}_{F}^{(2)} | {\cal W}_1^{2,3} \rangle^{ab}
+ \tilde Q_{B}^{(3)} | 
{\cal W}_{1}^{3,2} \rangle^{ab} -
{\tilde Q}_F^{(1)} | {\cal X}_1 \rangle^{ab} =0
\ee
\be \label{inv-w-2b}
  {
  \tilde Q}_{F}^{(1)} | {\cal W}_2^{1,3} \rangle^{ab}
+ \tilde Q_{B}^{(3)} | 
{\cal W}_{2}^{3,1} \rangle^{ab} -
{\tilde Q}_F^{(2)} | {\cal X}_2 \rangle^{ab} =0
\ee
\be \label{inv-w-3b}
 {\tilde Q}_{F}^{(1)}
| {\cal W}_3^{1,2} \rangle^{ab}
-
{\tilde Q}_{F}^{(2)}  
|{\cal W}_3^{2,1} \rangle^{ba}
-
\tilde Q_{B}^{(3)} 
|{\cal X}_{3} \rangle^{ab} =0
\ee
Note that the equations \eqref{xy-1b}-\eqref{inv-w-3b} should hold only when acting on fields and transformations satisfying the constraints of equations \eqref{gaugefixb}, \eqref{gaugefixf}, and \eqref{gaugefixf2}. The generalization to the unconstrained case was written in \cite{Buchbinder:2021qrg}.

\section{\texorpdfstring{$N=1$}{N=1} Super Yang-Mills} \setcounter{equation}0
\label{sectionsym}
Let us turn to particular examples.

For the case of Super Yang-Mills 
 one  introduces  colour indices
in the equations \p{LNSINT}--\p{GTNSINT} and in 
\p{int.}--\p{BFFG-2-2}  and takes the fields in the bosonic and the fermionic sectors
as
\be \label{YM-333}
| \Phi^{(i)} \rangle^A = ({\phi}_\mu^A(x) \alpha_2^{\mu (i), +}
-i { E}^A(x) c_0^{(i), + } b_2^{(i), + })|0^{(i)}  \rangle,
\ee
\be \label{psi-YM-1}
| \Psi^{(i)} \rangle^{a,A} =  
\Psi^{a,A} (x) |0^{(i)}\rangle
\ee
 the only non-zero parameter of gauge transformations being
\be
| \Lambda^{(i)} \rangle^A = i b_2^{(i),+} \lambda^A(x) | 0^{(i)} \rangle
\ee
The full interacting cubic Lagrangian is a sum of
\p{L-bbb-gf}
and of\footnote{Note that in Section \ref{subffb} we placed the boson in the third Fock space, whereas now we have three copies of the boson's Fock space in addition to three copies of the fermion's. To write the interaction in a symmetric way, we introduce a cyclic sum over the Fock space indices.}
\be\label{LNSINT-YM-2}
{ \cal L}_{\text{int}} =\sum_{i=1}^3 
{}^A\langle \Psi^{(i)} |g_{0}^{(i)}| \Psi^{(i)}
\rangle_A + 
g \left ( {}^A\langle \Psi^{(1)}| {}^B\langle \Psi^{(2)}| {}^C\langle \Phi^{(3)}| 
|{\cal V} \rangle_{ABC}  + \,\, \text{cyclic}  \right )
\ee
The  cubic interaction  vertex between three bosons
is given by the expression \p{sbv-3} with added  colour indices
\be \label{YM-vvv} 
| V \rangle_{ABC}= -\frac{i}{12} f_{ABC} 
{\cal Z}_{222}
 \times c_0^{(1)}c_0^{(2)}c_0^{(3)}
\,\, 
| 0^{(1)} \rangle \otimes
|0^{(2)} \rangle \otimes | 0^{(3)} \rangle  \,\, 
\ee
The cubic interaction vertex  between two fermions with spins one-half and one boson with spin one is
\be \label{YM-222}
|{\cal V} \rangle^{ab}_{ABC} = \frac{i}{3} f_{ABC}(\alpha_2^{(3),+} \cdot \gamma)^{ab} \,\, 
| 0^{(1)} \rangle \otimes
|0^{(2)} \rangle \otimes | 0^{(3)} \rangle + \,\, \text{cyclic}
\ee
The only non-trivial $| {\cal W} \rangle$ vertices in the solution of equations \p{xy-1b}--\p{xy-3b} are given by
\be \label{YM-444}
| {\cal W}_1^{2,3} \rangle^{ab}_{ABC} =
| {\cal W}_2^{1,3} \rangle^{ab}_{ABC}
= f_{ABC}\, C^{ab} c_2^+\,
 | 0^{(1)} \rangle \otimes
|0^{(2)} \rangle \otimes | 0^{(3)} \rangle
\ee
The vertex \p{YM-222} enters the Lagrangian and determines the interaction
between the vector field and two fermions, whereas the vertices \p{YM-444}
express the nonlinear part of the gauge transformations.

After eliminating  the auxiliary field  $E^A(x)$ via its own equations of motion one obtains an action for $N=1$ Super Yang-Mills 
up to the cubic order.
Alternatively,  one  could have imposed an off-shell transversality constraint
on the physical field $\phi_\mu^A(x)$ similarly to how it was
 done for the supergravity multiplets (see the discussion around
the equations \p{off-shell-1} -- \p{L-bbb-gf}).
This would have put an auxiliary field $E^A(x)$ equal to zero and restricted
the parameter of gauge transformations as $\Box \lambda(x) =0$. 

A higher spin generalization  is given in \cite{Buchbinder:2021qrg},
by multiplying the vertex \p{YM-222} with an arbitrary function
of the BRST invariant expressions \p{sbv-1-x}--\p{sbv-3}
and then finding corresponding $| {\cal W} \rangle$ and $| {\cal X} \rangle$
vertices. These solutions are covariant versions of the vertices 
found in the light-cone formalism \cite{Metsaev:2007rn}.

Because of the cubic interactions, the  supersymmetry transformations
for the fermion in \p{susyveclin}
will be deformed with a nonlinear term
\bea \label{SUSY-sym}
\delta^\prime | \Psi^{(i)} \rangle^{a}_A & = & g \,  f_{ABC}\, \times \\ \nonumber  
&\times& {}_B\langle \phi^{(i+1)}| {}_C\langle \phi^{(i+2)}|
(\gamma^{\mu \nu})^a{}_b\alpha^{(i+1),+}_{2,\mu}  \alpha^{(i+2),+}_{2,\nu} \epsilon^b \,\,
| 0^{(i+1)} \rangle \otimes
|0^{(i+2)} \rangle \otimes | 0^{(i)} \rangle
\eea
being the standard supersymmetry transformations for the $N=1$  Yang-Mills supermultiplet.
Let us notice, however, that if one imposes the off-shell transversality condition
on the vector field, the supersymmetry transformations put the fields completely on shell.
In this way one considers on-shell cubic vertices, which transform into each other
under linear supersymmetry transformations \p{susyveclin}.

\section{\texorpdfstring{$N=1$}{N=1} Supergravities} \setcounter{equation}0
\label{sectionsugra}
In the following we show how we can write the cubic interaction vertices of $N=1$ supergravity in ${\cal D}=4,6$ and $10$
dimensions. These can be compared with the full Lagrangians of the theories which we include in Appendix \ref{AppendixB}.
 
The cubic vertices for $N=1$ supergravities,
which describe a nonlinear deformations of the Lagrangian \p{ltot}
can be divided into two types.
The first type is universal 
in the same sense as
is  the Lagrangian \p{ltot}, i.e., 
the vertices have the same form in ${\cal D}=4,6$ and $10$.
The second type of the vertices are 
specific to particular dimensions.

In this section we consider the cubic interaction vertices of supergravities whose free versions were written in Section \ref{linsugra}. Namely, these are supergravity in ${\cal D}=4$ coupled to a chiral multiplet, supergravity in ${\cal D}=6$ coupled to a $(1,0)$ tensor multiplet, and pure supergravity in ${\cal D} = 10$.

In Subsection \ref{sugrauni} we write the ``universal'' vertices, which are present in ${\cal D}=4$, 6, and 10. In Subsections \ref{sugra10} and \ref{sugra6} we write specific boson-fermion-fermion vertices for ${\cal D}=10$ and 6, respectively.

\subsection{Universal vertices} \label{sugrauni}
Let us start with the first type of cubic vertices. 
We impose  off-shell transversality constraints
on the physical fields $\phi_{\mu,\nu}(x)$ and $\Psi_\mu^a(x)$.
Therefore, we have
\be
| \phi^{(i)} \rangle =  \phi_{\mu, \nu}(x) \,\alpha_{1}^{\mu (i),+} \alpha_2^{\nu (i),+} 
|0^{ (i)} \rangle, \qquad | \Psi^{(i)}\rangle^a = \Psi_\mu^{a}(x) \,  \alpha_1^{\mu (i),+} | 0^{ (i)} \rangle
\ee
We take the cubic vertices for three bosons 
\be \label{sugralike3b}
  -2\, g \, \langle \phi^{(1)} | \,\,
\langle \phi^{(2)}| \,\,
\langle \phi^{(3)}| {\cal Z}_{111}
{\cal Z}_{222}  
| 0^{(1)} \rangle \otimes
|0^{(2)} \rangle \otimes | 0^{(3)} \rangle
\ee
where the expressions for ${\cal Z}_{mnp}$ are given in
\p{sbv-3}.
For two fermions and one boson we take the vertex
\be \label{sugralikebff}
g \,\,    \langle \phi^{(3)} | \,\,
{}_a\langle \Psi^{(1)}| \,\,
{}_b\langle \Psi^{(2)}| {\cal Z}_{111}(\gamma \cdot \alpha_2^{(3),+} )^{ab}
| 0^{(1)} \rangle \otimes
|0^{(2)} \rangle \otimes | 0^{(3)} \rangle
+ \,\, \text{cyclic}.
\ee
which solves the equations \p{xy-1b}--\p{xy-3b} and \p{inv-w-1b}--\p{inv-w-3b}
with
\bea \label{s-w-ym-1}
({\cal W}_3^{1,2})^{ab}&=& \,c^{(2),+}_1
 ( \gamma \cdot \alpha_2^+)^{ab}
 (\alpha^{(3), +}_1 \cdot \alpha^{(1), +}_1) \\  \nonumber
({\cal W}_3^{2,1})^{ab}&=&  c^{(1),+}_1
  (\gamma \cdot \alpha_2^{+} )^{ab} 
 (\alpha^{(2), +}_1 \cdot \alpha^{(3), +}_1)
\\  \nonumber
({\cal W}_1^{3,2})^{ab}&=& -\,c^{(2),+}_1
 (p^{(1)} \cdot \gamma)^a{}_c
 (\gamma \cdot \alpha_2^{+} )^{cb}
 (\alpha^{(3), +}_1 \cdot \alpha^{(1), +}_1)
\\  \nonumber
({\cal W}_2^{3,1})^{ab}&=&  -c^{(1),+}_1
 (p^{(2)} \cdot \gamma)^a{}_c ( \gamma \cdot \alpha_2^{+})^{cb}
 (\alpha^{(2), +}_1 \cdot \alpha^{(3), +}_1)
\\  \nonumber
({\cal W}_1^{2,3})^{ab}&=& -c^+_2
  C^{ab} {\cal Z}_{111} +
  c_1^{(3),+}
(p^{(1)} \cdot \gamma)^a{}_c 
(\gamma \cdot \alpha_2^+)^{cb}
 (\alpha^{(1), +}_1 \cdot \alpha^{(2), +}_1)
   \\
  \nonumber
({\cal W}_2^{1,3})^{ab}&=& c_2^+ C^{ab}  {\cal Z}_{111}
+ 
c_1^{(3),+}
 (p^{(2)} \cdot \gamma)^a{}_c 
 (\gamma \cdot \alpha_2^+)^{cb}
  (\alpha^{(1), +}_1 \cdot \alpha^{(2), +}_1)
\eea
and
\bea \label{x-sol-1} \nonumber
{\cal X}_1^{ab}&=&  c_1^{(2),+} c_2^+ b_1^{(1),+} C^{ab}
 (p^{(1)} \cdot \alpha_1^{(3),+}) - 
 c_1^{(2),+} c_1^{(3),+} b_1^{(1),+}
(p^{(1)}\cdot \gamma)^a{}_c (\gamma \cdot \alpha_2^+)^{cb}
\\
  \nonumber
{\cal X}_2^{ab}&=& -c_2^+  c_1^{(1),+}  b_1^{(2),+} C^{ab}
 (p^{(2)} \cdot \alpha_1^{(3),+}) + 
 c_1^{(3),+} c_1^{(1),+} b_1^{(2),+}
(p^{(2)}\cdot \gamma)^a{}_c (\gamma \cdot \alpha_2^+)^{cb}
\\
{\cal X}_3^{ab}&=&-c_1^{(1),+} c_1^{(2),+}b_1^{(3)+}
(\gamma \cdot \alpha_2^+)^{ab}
\eea
The total Lagrangian, which is the sum of \p{L-bbb-gf} and
of
\be\label{LNSINT-SG-2}
{ \cal L}_{\text{int}} =\sum_{i=1}^3 
\langle \Psi^{(i)} |g_{0}^{(i)}| \Psi^{(i)}
\rangle + 
g \left ( \langle \Psi^{(1)}| \langle \Psi^{(2)}| \langle \Phi^{(3)}| 
|{\cal V} \rangle  + \,\, \text{cyclic}  \right )
\ee
has the form
\bea \label{t2}
{ \cal L}&=&  - \phi^{\mu, \nu} \Box \phi_{\mu,\nu }
+48g ( \partial_\rho \partial_\tau \phi_{\mu, \nu}) \phi^{\mu, \nu} \phi^{\rho, \tau }
- 96 g (\partial_\rho \partial_\tau \phi_{\mu, \nu}) \phi^{\mu, \tau} \phi^{\rho, \nu} \\ \nonumber
 &-&\frac12 {\bar\Psi}^\mu \gamma^\nu \partial_\nu \Psi_\mu + 12ig \phi^{\mu,\nu} {\bar \Psi}^\alpha \gamma_\nu \partial_\alpha \Psi_\mu - 6ig \phi^{\mu,\nu} {\bar \Psi}^\alpha \gamma_\nu \partial_\mu \Psi_\alpha
\eea
Let us note that the overall coefficients
in the cubic vertices \p{sugralike3b} and \p{sugralikebff}
are not fixed by the requirement
of the gauge invariance and this particular choice is dictated by the supersymmetry transformations.
However,  due to the off-shell transversality  conditions the supersymmetry transformations
\be
\phi_{ \nu, \mu }(x) = i\, \bar \Psi_{\mu}(x) \gamma_\nu \, \epsilon, \quad
\delta \Psi_\mu(x) = - \gamma^\nu\gamma^\rho\epsilon \,\partial_\nu \phi_{\rho,\mu}(x)
\ee
put the fields
completely on shell.

Similarly to the case of cubic vertices in Super Yang-Mills, the supergravity vertices of the first type 
\p{sugralikebff}
can be generalized
to higher spins \cite{Buchbinder:2021qrg} by multiplying them by an arbitrary function
of the BRST invariant expressions \p{sbv-1-x}--\p{sbv-3}
and
finding corresponding $| {\cal W} \rangle$ and $| {\cal X} \rangle$
vertices.

\subsection{Vertices of {${\cal D}=10$} supergravity} \label{sugra10}
In order to consider the cubic vertices of the second type it is easier
to decompose the fermionic fields into irreducible representations of Poincar\'e group 
  according to \p{dec-2}.
Then in ten dimensions we have the vertex
\be   \label{hs-vsugra-3}
  {}\langle \phi^{(3)} | \,\,
{}_a\langle \psi^{(1)}| \,\,
{}_b\langle \psi^{(2)}|  | {\cal V}_{{\cal J}} \rangle^{ab}, 
\ee
where
\bea \label{vertex-7}
{\cal V}^{ab}_{{\cal J}}& =& (\gamma_{\mu  \tau \sigma \lambda \nu})^{ab} \alpha_1^{\mu (1), +} 
\alpha_1^{\nu (2), +} p^{\tau, (3)} \alpha_1^{\sigma (3),+} \alpha_2^{\lambda (3),+} +
\\ \nonumber
&+& ( \alpha_1^{(3),+} \cdot \gamma)^{ab} 
[(p^{(3)} \cdot \alpha_1^{(1),+} )(\alpha_1^{(2),+} \cdot \alpha_2^{(3),+})
-
(p^{(3)} \cdot \alpha_1^{(2),+} )(\alpha_1^{(1),+} \cdot \alpha_2^{(3),+})]
- \\ \nonumber
&-& ( \alpha_2^{(3),+} \cdot \gamma)^{ab} 
[(p^{(3)} \cdot \alpha_1^{(1),+} )(\alpha_1^{(2),+} \cdot \alpha_1^{(3),+})
-
(p^{(3)} \cdot \alpha_1^{(2),+} )(\alpha_1^{(1),+} \cdot \alpha_1^{(3),+})]+ \\ \nonumber
&+& (p^{(3)} \cdot \gamma )^{ab}[(\alpha_1^{(1),+} \cdot \alpha_2^{(3),+})
(\alpha_1^{(2),+} \cdot \alpha_1^{(3),+})
-
(\alpha_1^{(2),+} \cdot \alpha_2^{(3),+})
(\alpha_1^{(1),+} \cdot \alpha_1^{(3),+})]
\eea
This vertex corresponds to the coupling of the field $B_{\mu \nu}(x)$
with two gravitini.
The corresponding ${\cal W}$ vertices have the form
 \bea 
({\cal W}_3^{2,1})^{ab}&=&   \frac12 c^{(1),+}_1 \alpha_1^{\nu(2),+} \alpha_1^{\sigma(3),+} \alpha_2^{\lambda(3),+} (\gamma_{\sigma\lambda}\gamma_{\nu})^{ab} \\
    \nonumber
({\cal W}_3^{1,2})^{ab}&=&   \frac12 c^{(2),+}_1 \alpha_1^{\nu(1),+} \alpha_1^{\sigma(3),+} \alpha_2^{\lambda(3),+} (\gamma_{\sigma\lambda}\gamma_{\nu})^{ab} \\
    \nonumber    
({\cal W}_2^{3,1})^{ab}&=& -c^{(1),+}_1 [\frac12 (p^{(2)}\cdot \gamma)^a{}_c \alpha_1^{\nu(2),+} \alpha_1^{\sigma(3),+} \alpha_2^{\lambda(3),+} (\gamma_{\sigma\lambda}\gamma_{\nu})^{cb} + \\ \nonumber
&\quad&\quad+ p^{(1),\mu} \alpha_1^{\nu(2),+} \alpha_1^{\sigma(3),+} \alpha_2^{\lambda(3),+}(\eta_{\mu\nu}\gamma_\lambda\gamma_\nu - \eta_{\mu\lambda}\gamma_\sigma\gamma_\nu + \eta_{\mu\nu}\gamma_{\sigma\lambda})^{ab}] \\ \nonumber    
({\cal W}_1^{3,2})^{ab}&=& c^{(2),+}_1 [\frac12 (p^{(2)}\cdot \gamma)^b{}_c \alpha_1^{\nu(1),+} \alpha_1^{\sigma(3),+} \alpha_2^{\lambda(3),+} (\gamma_{\sigma\lambda}\gamma_{\nu})^{ca} + \\ \nonumber
&\quad&\quad+ p^{(2),\mu} \alpha_1^{\nu(1),+} \alpha_1^{\sigma(3),+} \alpha_2^{\lambda(3),+}(\eta_{\mu\nu}\gamma_\lambda\gamma_\nu - \eta_{\mu\lambda}\gamma_\sigma\gamma_\nu + \eta_{\mu\nu}\gamma_{\sigma\lambda})^{ba}]
\eea
The solution for the group structure equations includes
\bea \nonumber
{\cal X}_1^{ab} &=& -b_1^{(1),+} [c_1^{(3),+}c_1^{(2),+}(\gamma\cdot \alpha_2^{(3),+})^a{}_c (\gamma\cdot p^{(1)})^{cb} - c_2^{(3),+}c_1^{(2),+}(\gamma\cdot \alpha_1^{(3),+})^a{}_c (\gamma\cdot p^{(1)})^{cb}] \\ \nonumber
{\cal X}_2^{ab} &=& b_1^{(2),+} [c_1^{(3),+}c_1^{(1),+}(\gamma\cdot \alpha_2^{(3),+})^a{}_c (\gamma\cdot p^{(2)})^{cb} - c_2^{(3),+}c_1^{(1),+}(\gamma\cdot \alpha_1^{(3),+})^a{}_c (\gamma\cdot p^{(2)})^{cb}] \\
{\cal X}_3^{ab} &=& b_1^{(3),+} c_1^{(3),+}c_1^{(1),+}(\gamma\cdot \alpha_2^{(3),+})^{ab} - b_2^{(3),+} c_2^{(3),+}c_1^{(1),+}(\gamma\cdot \alpha_1^{(3),+})^{ab}
\eea

The remaining cubic vertices in ${\cal D}=10$ are those which correspond to the coupling of the $B_{\mu \nu}(x)$ to one gravitino and one dilatino,
\be  \label{vertex-8}
  {}\langle \phi^{(3)} | \,\,
{}_a\langle \psi^{(1)}| \,\,
{}^b\langle \Xi^{(2)}|  | {\cal V}_{{\cal L}} \rangle^{a}_{b}, 
\ee
There are two such couplings. The first is
\bea \label{vertex-L1}
({\cal V}_{{\cal L}_1})^a{}_b  =
 (\gamma_{\mu  \nu  \tau})^{ac}  
 p^{\mu, (3)} \alpha_1^{\nu (3),+} \alpha_2^{\tau (3),+}  (\alpha_1^{ (1), +} \cdot \gamma )_{cb} 
\eea
The non-trivial ${\cal W}$ vertices are
\bea 
({\cal W}_3^{2,1})^{ab}&=&   -c^{(1),+}_1 (\gamma_{\nu\tau})^{ab} \alpha_1^{\nu(3),+}\alpha_2^{\tau(3),+}
 \\ \nonumber
   ({\cal W}_2^{3,1})^{ab}&=& c^{(1),+}_1 \big[ (\gamma_{\nu\tau})^{a}{}_c \alpha_1^{\nu(3),+}\alpha_2^{\tau(3),+} (p_2\cdot \gamma)^{cb}\big]
\eea
The second vertex of this type is
\bea \label{vertex-L2}
({\cal V}_{{\cal L}_2})^a{}_b  &=& \delta^a{}_b  
  (\alpha_1^{ (3),+} \cdot \alpha_2^{ (3),+})  (\alpha_1^{ (1), +} \cdot  p^{ (3)})  
\eea
for which
\bea 
({\cal W}_3^{2,1})^{ab}&=&   
 -\frac12 c^{(1),+}_1 C^{ab}  \alpha_1^{(3),+}\cdot\alpha_2^{(3),+}
 \\ \nonumber
   ({\cal W}_2^{3,1})^{ab}&=& \frac12 c^{(1),+}_1 (p_2\cdot \gamma)^{ab} \alpha_1^{(3),+}\cdot\alpha_2^{(3),+} 
\eea
For the last two vertices the solutions for the group structure equations are with $|{\cal X}_i\rangle = 0$.

\subsection{Vertices of ${\cal D}=6$ supergravity} \label{sugra6}
Most of the vertices of ${\cal D}=6$ supergravity have already been described above. They include the universal vertices of Subsection \ref{sugrauni}, as well as the vertices \eqref{vertex-L1} and \eqref{vertex-L2}, which are both present in ${\cal D}=6$.

The vertex that has a different form is the coupling of the $B$-field to two gravitini, which is now given by
\be {\cal V}^{ab} = (\gamma_\tau \gamma_{\mu\nu\rho}\gamma_{\lambda})^{ab} \alpha_1^{\lambda(1),+}\alpha_1^{\tau(2),+}p^{\mu(3)}\alpha_1^{\mu(3),+}\alpha_2^{\nu(3),+} \ee
The corresponding ${\cal W}$ vertices have the form
 \bea 
({\cal W}_3^{2,1})^{ab}&=& c_1^{(1),+}\alpha_2^{\mu(2),+}\alpha_1^{\nu(3),+}\alpha_1^{\rho(3),+}(\gamma_{\nu\rho}\gamma_\mu)^{ab}\\
    \nonumber
({\cal W}_3^{1,2})^{ab}&=&  c_1^{(2),+}\alpha_1^{\mu(2),+}\alpha_1^{\nu(3),+}\alpha_1^{\rho(3),+}(\gamma_{\nu\rho}\gamma_\mu)^{ab}\\
    \nonumber    
({\cal W}_2^{3,1})^{ab}&=& -c_1^{(1),+}\big[ \alpha_1^{\nu(3),+}\alpha_2^{\rho(3),+} (\gamma_{\nu\rho})^a{}_c p^{\mu(2)}\alpha_1^{\sigma(2),+}(\gamma_\mu\gamma_\sigma)^{cb} + \\ \nonumber
&\quad&\qquad\qquad +2(p^{(1)}\cdot \alpha_1^{(2),+})\alpha_1^{\nu(3),+}\alpha_2^{\rho(3),+} (\gamma_{\nu\rho})^{ab}\big]\\ \nonumber   
({\cal W}_1^{3,2})^{ab}&=& c_1^{(2),+}\big[ \alpha_1^{\nu(3),+}\alpha_2^{\rho(3),+} (\gamma_{\nu\rho})^a{}_c p^{\mu(1)}\alpha_1^{\sigma(1),+}(\gamma_\mu\gamma_\sigma)^{cb} + \\ \nonumber
&\quad&\qquad\qquad +2(p^{(2)}\cdot \alpha_1^{(1),+})\alpha_1^{\nu(3),+}\alpha_2^{\rho(3),+} (\gamma_{\nu\rho})^{ab}\big]
\eea
with
\bea \nonumber
{\cal X}_1^{ab} &=& -b_1^{(1),+} [c_1^{(3),+}c_1^{(2),+}(\gamma\cdot \alpha_2^{(3),+})^a{}_c (\gamma\cdot p^{(1)})^{cb} - c_2^{(3),+}c_1^{(2),+}(\gamma\cdot \alpha_1^{(3),+})^a{}_c (\gamma\cdot p^{(1)})^{cb}] \\ \nonumber
{\cal X}_2^{ab} &=& b_1^{(2),+} [c_1^{(3),+}c_1^{(1),+}(\gamma\cdot \alpha_2^{(3),+})^a{}_c (\gamma\cdot p^{(2)})^{cb} - c_2^{(3),+}c_1^{(1),+}(\gamma\cdot \alpha_1^{(3),+})^a{}_c (\gamma\cdot p^{(2)})^{cb}] \\
{\cal X}_3^{ab} &=& 2b_1^{(3),+} c_1^{(1),+}c_1^{(2),+}(\gamma\cdot \alpha_2^{(3),+})^{ab} - 2b_2^{(3),+} c_1^{(1),+}c_1^{(2),+}(\gamma\cdot \alpha_1^{(3),+})^{ab}
\eea
solving the group structure equations.

There is also the coupling of \(B_{\mu\nu}\) to two dilatini,
\be  \label{vertex-8-1}
  {}\langle \phi^{(3)} | \,\,
{}^a\langle \Xi^{(1)}| \,\,
{}^b\langle \Xi^{(2)}|  | {\cal V}_{{\cal X}} \rangle_{ab},
\ee
which is of the form
\be ({\cal V}_{\cal X})_{ab} = (\gamma_{\mu\nu\rho})_{ab}p^{\mu(3)}\alpha_1^{\nu(3),+}\alpha_2^{\rho(3),+} \ee
In this case all the ${\cal W}$ vertices are trivial.

\section{Light Cone Formalism}\label{sectionlightcone} \setcounter{equation}0
In this Section we describe how to construct cubic vertices of ${\cal D}=4$
$N=1$ super Yang-Mills and Supergravity in the light cone approach following \cite{Metsaev:2019dqt} 
(see also \cite{Brink:2005wh}--\cite{Ponomarev:2016lrm} for a brief review of the light cone approach).

\subsection{Set Up}
To construct cubic interaction vertices in the light cone approach 
let us consider a field theoretic realization
of the ${\cal D}=4$ $N=1$ super Poincar\'e algebra
\begin{align} \label{SPA}
[Q^a,Q^b]&= \frac{1}{2}(\gamma^{\mu})^{ab} P_\mu \,,\\
[Q^a,J^{\mu \nu}]&= \frac{1}{2}(\gamma^{\mu \nu})^a{}_b Q^b \,,\\
[J^{\mu \nu},P^\rho]&=P^\mu\eta^{\nu \rho}-P^\nu\eta^{\mu \rho}\,,\\
[J^{\mu \nu},J^{\rho \sigma}]&=J^{\mu \sigma}\eta^{\nu \rho}-J^{\nu \sigma}\eta^{\mu \rho}-J^{\mu \rho}\eta^{\nu \sigma}+J^{\nu \rho}\eta^{\mu \sigma}\,.
\end{align}
Here   $J^{\mu \nu}$ are
generators of Lorentz transformations, 
 $P^\mu$ are generators of translations, and 
 $Q^a$ are generators of Supersymmetry transformations.
 These generators are split into kinematical and dynamical generators.
 Kinematical generators preserve the Cauchy surface (the light cone) and are quadratic in fields
 both on free and interacting levels. The other generators are dynamical and they receive
 higher order corrections in fields. These corrections are determined from the requirement
 that the Poincar\'e algebra is preserved at the interacting level.
  
We choose the four dimensional  coordinates as
  \be
x^\pm = \frac{1}{\sqrt 2} (x^3 \pm x^0), \quad z = \frac{1}{\sqrt 2} (x^1 + i x^2), \quad \bar z = \frac{1}{\sqrt 2} (x^1 - i x^2)
\ee
The coordinate $x^+$ is treated as the time direction and $H=P^-$ is the Hamiltonian\footnote{ Note that $\beta$ is used instead of  $p^+$ 
in order to simplify the form of the equations.We shall also put
$x^+=0$ for now on.}. The generators of the super Poincar\'e algebra are  split according to
\begin{align}
\text{kinematical}&: && P^{+}, P^z, { P}^{\bar z}, J^{z+}, { J}^{\bar z +}, J^{+-}, J^{z \bar z}, Q^{+},{\bar Q}^{+}, &&: 9\\
\text{dynamical}&: && P^{-}, J^{z-}, { J}^{\bar z-}, Q^{-},{\bar Q}^{-}   &&:5
\end{align}

It is sufficient to construct the Poincare algebra at $x^+=0$ and then evolve all the generators according to $\dot{G}=i[H,G]$. The  equations to be solved are
\be\label{hardequations}
[Q^{-},P^{-}]=[\bar Q^{-},P^{-}]=0\,, \quad
[J^{z,-},P^{-}]=[J^{\bar z,-},P^{-}]=0\,.
\ee
The spectrum consists of bosonic $\phi_\lambda(x)$ and fermionic
$\psi_\lambda(x)$ fields\footnote{In this Section, unlike the previous ones, the index $\lambda$ denotes a helicity of
a field, rather than its Lorentz index.}
with the helicities $\lambda= \pm 1, \pm 2$ for bosons,
$\lambda= \pm \frac{1}{2}, \pm \frac{3}{2}$ for fermions.
It is convenient to work with partial Fourier transforms
\begin{align}
    \phi_\lambda(x)&=(2\pi)^{-\tfrac{3}2} \int e^{+i(x^-\beta +z \bar p + \bar z p)} \phi_\lambda(\vec{p})\, d^{3}p\,,\\
    \psi_\lambda(x)&=(2\pi)^{-\tfrac{3}2} \int e^{+i(x^-\beta + z \bar p + \bar z p)} \psi_\lambda(\vec{p})\, d^{3}p\,
\end{align}
with $d^{3}p= d \beta \, d p \, d {\bar p}$.
The fields obey the following conjugation rules
\be
\phi^\dagger_\lambda(\vec{p}) = \phi_{-\lambda}(-\vec{p}), \quad \psi^\dagger_\lambda(\vec{p}) = \psi_{-\lambda}(-\vec{p})
\ee
Introducing a Grassman momentum $p_\theta$, one can combine 
the bosonic and fermionic fields into superfields
\be
\Phi_\lambda = \phi_\lambda + \frac{p_\theta}{\beta}\psi_{\lambda -\frac{1}{2}}, 
\quad
\Phi_{-\lambda + \frac{1}{2}} = \psi_{-\lambda + \frac{1}{2}} + {p_\theta}\phi_{-\lambda}, 
\ee
with conjugation properties
\be
\overline \Phi_{-\lambda} =  \phi_{-\lambda}^\dagger + \frac{p_\theta}{\beta}\psi^\dagger_{-\lambda +\frac{1}{2}}, 
\quad
\overline \Phi_{\lambda - \frac{1}{2}} = -\psi_{\lambda - \frac{1}{2}}^\dagger
+ {p_\theta}\phi_{\lambda}^\dagger, 
\ee
 The equal time Poisson brackets between the fields
\be\label{equaltime1}
    [\phi_{\lambda}(\vec{p}),\phi_{\lambda^\prime}^\dagger(\vec{p}^\prime)]=\delta_{\lambda,\lambda^\prime}\frac{\delta^{3}(\vec{p}-\vec{p}^\prime)}{2\beta},
 \quad
 [\psi_{\lambda}(\vec{p}),\psi_{\lambda^\prime}^\dagger(\vec{p}^\prime)]=\delta_{\lambda,\lambda^\prime}\frac{\delta^{3}(\vec{p}-\vec{p}^\prime)}{2}
\ee
 read in terms of the superfields as
\be\label{equaltime3}
    [\Phi_{\lambda}(\vec{p},p_\theta),\overline \Phi_{\lambda^\prime}(\vec{p}^\prime,p_\theta^\prime)]=(-)^{\epsilon_{\lambda+\frac{1}{2}}}\delta_{\lambda,\lambda^\prime+\frac{1}{2}}
    \frac{\delta^{3}(\vec{p}-\vec{p}^\prime)\,
    \delta(p_\theta-p_\theta^\prime)}
    {2\beta} 
 \ee
 where $\epsilon_\lambda$ is $0$ for integer $\lambda$ and is $1$ for half-integer $\lambda$.

The kinematical generators, which are the same both on free and interacting levels  have the form
\be
{P}^+=\beta\,, \quad {P}^z=p, \quad P^{\bar z}=\pb,
\quad
{J}^{z+}=-\beta\pfrac{\pb}, \quad  {J}^{\zb+}=-\beta\pfrac{p},
\ee
\be \nonumber
 {J}^{-+}=-\frac{\pl}{\pl \beta} \beta - \frac{1}{2}p_\theta \frac{\partial}{\partial p_{\theta}} + \frac{1}{2} \epsilon_\lambda,
\quad {J}^{z\zb}= p\pl_p-\bar p\frac{\partial}{\partial \bar p} +\lambda -p_\theta \frac{\partial}{\partial p_{\theta}} 
\ee
\be \nonumber
Q^+ = (-)^{\epsilon_\lambda} \beta \frac{\partial}{\partial p_\theta}, \quad \bar Q^{+}= (-)^{\epsilon_\lambda} p_\theta
\ee
  The dynamical generators at the free level are
\bea
    H_2&=&-\frac{p\pb}{\beta}\,,  \\ \nonumber
        {J}^{z-}_2&=& -\pfrac{\pb} \frac{ p\pb}{\beta} +p \pfrac{\beta}  \nonumber
        - \left (\lambda - \frac{1}{2} p_\theta \frac{\partial}{\partial p_\theta} \right ) \frac{p}{\beta} 
        + \left( \frac{1}{2} p_\theta\frac{\partial}{\partial p_\theta}-\frac{1}{2}\epsilon_\lambda \right) \frac{ p}{\beta}
        \\ \nonumber
         {J}^{\zb-}_2&=&- \pfrac{p} \frac{ p\pb}{\beta} +\pb \pfrac{\beta}  
          + \left (\lambda - \frac{1}{2} p_\theta \frac{\partial}{\partial p_\theta} \right ) \frac{\bar p}{\beta} 
        + \left( \frac{1}{2} p_\theta\frac{\partial}{\partial p_\theta}-\frac{1}{2}\epsilon_\lambda \right) \frac{ p}{\beta} \\    \nonumber
    Q_2^- &=&(-)^{\epsilon_\lambda}\frac{p}{\beta} p_\theta \\ \nonumber
    \bar Q_2^- &=&(-)^{\epsilon_\lambda} \bar p \frac{\partial}{\partial p_\theta}
    \eea
    At the level of cubic interactions  one assumes the following expansion for the dynamical generators
\bea \label{cubiclc}
    H_3&=&H_2+\int d \Gamma_{[3]} \, 
    \overline \Phi^{\lambda_1 \lambda_2 \lambda_3}_{q_1 q_2 q_3} \,
    h_{\lambda_1 \lambda_2 \lambda_3}^{q_1 q_2 q_3}
    \\ \nonumber
     Q_3^{-}&=&Q_2^{-}
    +\int d \Gamma_{[3]} \,
    \overline \Phi^{\lambda_1 \lambda_2 \lambda_3}_{q_1 q_2 q_3}\,
    q_{\lambda_1 \lambda_2 \lambda_3}^{q_1 q_2 q_3} \\ \nonumber
    {\bar Q}_3^{-}&=&{\bar Q}_2^{-}
    +\int d \Gamma_{[3]}\, 
    \overline \Phi^{\lambda_1 \lambda_2 \lambda_3}_{q_1 q_2 q_3} \,
    {\bar q}_{\lambda_1 \lambda_2 \lambda_3}^{q_1 q_2 q_3} \\ \nonumber
    J^{z-}_3&=&J^{z-}_2+ \int d \Gamma_{[3]} \times \\ \nonumber
    &\times&\left[ 
    \overline \Phi^{\lambda_1 \lambda_2 \lambda_3}_{q_1 q_2 q_3} \,\,
    j_{\lambda_1 \lambda_2 \lambda_3}^{q_1 q_2 q_3}-
    \frac{1}{3}
    \left(\sum_{k=1}^3 \frac{\partial 
    \overline \Phi^{\lambda_1 \lambda_2 \lambda_3}_{q_1 q_2 q_3}
    }{\partial \bar{q}_k}\right)h_{\lambda_1 \lambda_2\lambda_3}^{q_1 q_2 q_3}-
     \frac{1}{3}
     \left(\sum_{k=1}^3 
     \frac{\partial
     \overline \Phi^{\lambda_1 \lambda_2 \lambda_3}_{q_1 q_2 q_3}
     }{\partial {q}_{\theta, k}}\right) q_{\lambda_1 \lambda_2\lambda_3}^{q_1 q_2 q_3}
    \right]\, 
    \\ \nonumber
    J^{\zb-}_3&=&J^{\bar z-}_2+ \int d \Gamma_{[3]} \times \\ \nonumber
    &\times &\left[
    \overline \Phi^{\lambda_1 \lambda_2 \lambda_3}_{q_1 q_2 q_3} \,\,
    \jb_{\lambda_1 \lambda_2\lambda_3}^{q_1 q_2 q_3}-
    \frac{1}{3} 
    \left(\sum_{k=1}^3 \frac{\partial 
    \overline \Phi^{\lambda_1 \lambda_2 \lambda_3}_{q_1 q_2 q_3}
    }{\partial q_k}\right) h_{\lambda_1 \lambda_2\lambda_3}^{q_1 q_2 q_3}
   + \frac{1}{3}
   \left(\sum_{k=1}^3 \frac{q_{\theta,k}}{\beta_k}\right) \overline \Phi^{\lambda_1 \lambda_2 \lambda_3}_{q_1 q_2 q_3}
   {\bar q}_{\lambda_1 \lambda_2\lambda_3}^{q_1 q_2 q_3}
  \right  ] 
   \eea
where  $\overline \Phi^{\lambda_1 \lambda_2 \lambda_3}_{q_1 q_2 q_3} \equiv
\overline \Phi^{\lambda_1}_{q_1}\overline \Phi^{\lambda_2}_{q_2}
  \overline  \Phi^{\lambda_3}_{q_3}$ and
\be
d \Gamma_{[3]}= (2 \pi)^3  \prod_{k=1}^3 \frac{d^3 q_k}{(2 \pi)^\frac{3}{2}} 
\delta^3\left(\sum_{i=1}^3 q_i\right)
\prod_{l=1}^3 dq_{ \theta,l }\,\delta\left(\sum_{j=1}^3 q_{\theta, j}\right)
\ee
is an integration measure.

\subsection{${\cal D}=4$, $N=1$ Super Yang Mills and Pure Supergravity}
The cubic vertices 
which are present in the interaction part of dynamical generators \p{cubiclc}
are determined from the requirement of preservation of the
algebra \p{SPA}.
A  solution 
 which contains the $N=1$ Super Yang Mills and Supergravity vertices 
has the form \cite{Metsaev:2019dqt}
\be \label{lcs1}
h_{\lambda_1 \lambda_2 \lambda_3}^{q_1 q_2 q_3} = C^{\lambda_1 \lambda_2 \lambda_3} 
(\PPb)^{M_\lambda +1} 
\prod_{i=1}^3
\beta_i^{-\lambda_i - \frac{1}{2}\epsilon_{\lambda_i}}
+
\overline C^{\lambda_1 \lambda_2 \lambda_3} 
(\PP)^{-M_\lambda -\frac{1}{2}} \PP_\theta
\prod_{i=1}^3
\beta_i^{\lambda_i - \frac{1}{2}\epsilon_{\lambda_i}}
\ee
\be \label{lcs2}
q_{\lambda_1 \lambda_2 \lambda_3}^{q_1 q_2 q_3} =- C^{\lambda_1 \lambda_2 \lambda_3} 
(\PPb)^{M_\lambda } \PP_\theta
\prod_{i=1}^3
\beta_i^{-\lambda_i - \frac{1}{2}\epsilon_{\lambda_i}}
\ee
\be \label{lcs3}
j_{\lambda_1 \lambda_2 \lambda_3}^{q_1 q_2 q_3} =2 C^{\lambda_1 \lambda_2 \lambda_3} 
(\PPb)^{M_\lambda }  \, \chi
\prod_{i=1}^3
\beta_i^{-\lambda_i - \frac{1}{2}\epsilon_{\lambda_i}}
\ee
and
\be
M_\lambda= \lambda_1 + \lambda_2 + \lambda_3, \quad \lambda_1= s_1 - \frac{1}{2}, \quad \lambda_2= s_2 - \frac{1}{2},
\quad
\lambda_3= - s_3 
\ee
In these equations  $C^{\lambda_1 \lambda_2 \lambda_3}$,
$\overline C^{\lambda_1 \lambda_2 \lambda_3}$
are coupling constants and
\be
    \PP=
    \frac13\left[ (\beta_1-\beta_2)p_3+(\beta_2-\beta_3)p_1+(\beta_3-\beta_1)p_2\right]   \,,\\
    \ee
\be
    \PP_\theta=
    \frac13\left[ (\beta_1-\beta_2)p_{\theta, 3}+(\beta_2-\beta_3)p_{\theta, 1}+(\beta_3-\beta_1)p_{\theta, 2}\right]   \,,\\
    \ee
\be
    \chi=\beta_1(\lambda_2-\lambda_3)+\beta_2(\lambda_3-\lambda_1)+\beta_3(\lambda_1-\lambda_2)\,.
\ee
The momenta $\beta_i, p_i, \bar p_i$ and $p_{\theta,i}$ obey the conservation properties as in \p{mcon}.
The vertices 
${\bar q}_{\lambda_1 \lambda_2 \lambda_3}^{q_1 q_2 q_3}$ and ${\bar j}_{\lambda_1 \lambda_2 \lambda_3}^{q_1 q_2 q_3}$
can be obtained from \p{lcs2}--\p{lcs3} by relevant hermitean conjugation.

The cubic vertices for $N=1$ Super Yang Mills  can be recovered by choosing
$s_1= s_2=s_3=1$ in the equations above. Similarly, pure $N=1$ Supergravity
vertices can be recovered by putting $s_1= s_2=s_3=2$.

\vskip 0.5cm

\noindent {\bf Acknowledgments.} M.T. would like to thank the organizers of the online conference ``Quarks 2020" 
and of the online workshop``Aspects of Symmetry" for the invitation to give talks. 
M.T. would like to thank the Department of Mathematics, the University of Auckland for the hospitality
during the final stage of the project. 
D.W. would like to thank the Yukawa Institute of Theoretical Physics (YITP) for its support in the period when this work was completed.
The work of I.L.B. and V.A.K. was partially supported by the Ministry of Education of Russian Federation, project FEWF-2020-0003.  The work of M.T. and D.W. was supported by the Quantum Gravity Unit
of the Okinawa Institute of Science and Technology Graduate University
(OIST). 

\renewcommand{\thesection}{A}

\renewcommand{\theequation}{A.\arabic{equation}}

\setcounter{equation}0
\appendix
\numberwithin{equation}{section}

\section{Conventions}\label{Appendix A}
We mainly  follow the notations of \cite{VanProeyen:1999ni}.

The Latin letters $a,b \ldots$ label spinorial indices.  The Greek letters
$\mu, \nu, \ldots$ label flat space-time vector indices
and  Greek letters with ``hat'' ${\hat \mu}, {\hat \nu},\ldots$ label vector indices in curved space-time.

We choose a real representation for Majorana spinors
\be
(\lambda^a)^\star=\lambda^a, \quad \bar \lambda_a =\lambda^b C_{ba}
\ee
The spinor indices can be raised and lowered by anti-symmetric charge conjugation matrices $C_{ab}$ and $C^{ab}$  as
\be
\lambda^a = C^{ab} \lambda_b, \quad \lambda_a = \lambda^b C_{ b a},
\quad C^{ab}C_{bc}=-\delta^a_c.
\ee
The $\gamma$--matrices satisfy the following anti-commutation relations
\begin{equation}\label{1}
(\gamma^\mu)^a{}_c (\gamma^\nu)^c{}_b
+
(\gamma^\nu)^a{}_c (\gamma^\mu)^c{}_b
 = 2 \eta^{\mu \nu} \delta^a_b.
\end{equation}
 In ${\cal D}=4$ the matrices $\gamma_\mu$ and $\gamma_{\mu \nu}$ with both spinorial indices  up (down)
 are symmetric and the matrices $C$, $\gamma_5$ and $\gamma_5 \gamma_\mu$ are anisymmetric. In ${\cal D}=10$ the matrices
 $\gamma_\mu$ and $\gamma_{\mu_1,...,\mu_5}$ with both spinorial indices up (down)
 are symmetric, and the matrices $\gamma_{\mu_1 \mu_2 \mu_3}$ are antisymmetric.

For checking the on-shell closure of the supersymmetry algebra and of the 
supersymmetry of the vertices we have used the following gamma-matrix identities
\be\label{FI}
(\gamma^\nu)_{ab}{(\gamma_ \nu)}_{ cd}+ (\gamma^\nu)_{ac}{(\gamma_ \nu)}_{ db} + (\gamma^\nu)_{ad}{(\gamma_ \nu)}_{ bc}=0,
\ee
\be
\gamma^\mu \gamma^{\nu_1, \nu_2, ... \nu_r} \gamma_\mu=
(-1)^r (D-2r) \gamma^{\nu_1, \nu_2,...\nu_r}.
\ee
For a product of gamma matrices we have
\be
\gamma^{\nu_1,..., \nu_i} \gamma_{\mu_1,..., \mu_j}=
\sum_{k =0}^{k =min (i,j)} \frac{i! j!}{(i-k)! (j-k)! k!} 
\gamma^{[\nu_1,...,\nu_{i-k}}{}_{[\mu_{k+1},...,\mu_{j}} \delta^{\nu_i}_{\mu_1} \delta^{\nu_{i-1}}_{\mu_2}...
\delta^{\nu_{n-k+1}]}_{\mu_k]}
\ee
and in particular
\be \label{GG-1}
\gamma_{\mu}\gamma_{\nu_1,\ldots \nu_k} = \gamma_{\mu,\nu_1,\ldots,\nu_k} + k \eta_{\mu[\nu_1}\gamma_{\nu_2,\ldots,\nu_k]} \ee

\section{$N=1$ Supergravities in ${\cal D}=4,6,10$} \label{AppendixB}\setcounter{equation}0  

The Lagrangian for ${\cal D}=10$ $N=1$  Supergravity is \cite{Chamseddine:1980cp} -- \cite{Bergshoeff:1981um}
     \begin{eqnarray} \nonumber
    L&=& - \frac{1}{2} R - \frac{1}{2} \bar { \psi}_{\hat \mu} \gamma^{\hat \mu \hat \nu \hat \rho} {\cal D}_{\hat \nu} 
    { \psi}_{\hat \rho} 
    - \frac{3}{4} \phi^{-\frac{3}{2}}H^{\hat \mu \hat \nu \hat \rho} H_{\hat \mu \hat \nu \hat \rho}
    -\\ \nonumber
    &-& \frac{1}{2} \bar \Xi \gamma^{\hat \mu} {\cal D}_{\hat \mu} \Xi - \frac{9}{16} \frac{\partial^{\hat \mu} \phi \, \partial_{ \hat \mu} \phi}{\phi^2}
    - \frac{3 \sqrt{2}}{8}\bar { \psi}_{\hat \mu} \gamma^{ \hat \nu} \gamma^{ \hat \mu} \Xi \, \frac{\partial_{ \hat \nu} \phi }{\phi} + \\ \nonumber
    &+& \frac{\sqrt{2}}{16} \phi^{-\frac{3}{4}} H_{\hat \nu  \hat \rho \hat \tau} 
    (\bar { \psi}_{ \hat \mu} \gamma^{\hat \mu \hat \nu \hat \rho \hat \tau \hat \lambda}  { \psi}_{\hat \lambda}
    + 6 \bar { \psi}^{ \hat \nu} \gamma^{\hat \rho }  { \psi}^{ \hat \tau} - \sqrt{2} 
    \bar { \psi}_{ \hat \mu} \gamma^{\hat \nu \hat \rho \hat \tau} \gamma^{ \hat \mu} \Xi) + \\ \nonumber
    &+&(\textrm{fermion})^4
    \end{eqnarray}
    where $H_{\hat \mu \hat \nu \hat \rho}= \partial_{[\hat \mu}B_{\hat \nu \hat \rho]}$ and ${\cal D}_{\hat \mu}$ is a covariant derivative
\be
{\cal D}_{{\hat \mu}} 
\Psi_{{\hat \nu}}^a =
\partial_{{\hat \mu}} \Psi_{{\hat \nu}}^a + 
\frac{1}{4} \omega_{{\hat \mu}}{}^{\rho \sigma} (\gamma_{\rho \sigma}
\Psi_{{\hat \nu}})^a
\ee
    The supersymmetry transformations with a local parameter $\epsilon^a(x)$ are
    \bea
    && \delta e_{\hat \mu}^\mu = \frac{1}{2}{\bar \epsilon} \, \gamma^{  \mu} \, \psi_{\hat \mu} \\ \nonumber
    && \delta \phi = -\frac{\sqrt 2}{3} \phi \, {\bar \epsilon} \, \Xi \\ \nonumber
    && \delta B_{\hat \mu \hat \nu} = \frac{\sqrt 2}{4} \phi^\frac{3}{4}\,({\bar \epsilon} \, \gamma_{\hat \mu} \, \psi_{\hat \nu}
    - {\bar \epsilon} \, \gamma_{\hat \nu} \, \psi_{\hat \mu} 
    -\frac{\sqrt 2}{2} \, {\bar \epsilon} \, \gamma_{\hat \mu \hat \nu} \Xi) \\ \nonumber
    && \delta \psi_{\hat \mu} =\left ( {\cal D}_{\hat \mu} + \frac{\sqrt 2}{32} \phi^{-\frac{3}{4}} (\gamma_{\hat \mu }{}^{\hat \nu \hat \rho \hat \sigma}
    - 9 \, \delta_{\hat \mu }^{\hat \nu} \, \gamma^{\hat \rho \hat \sigma})  \, H_{\hat \nu \hat \rho \hat \sigma} 
    \right )  \epsilon
    + (\textrm{fermion})^2 \\
    && \delta \Xi=-\frac{3 \sqrt{2}}{8} \phi^{-1} \, \gamma^{\hat \mu} \,  \partial_{\hat \mu} \phi \, \epsilon + \frac{1}{8} \phi^{-\frac{3}{4}} \,
    \gamma^{\hat \mu \hat \nu \hat \rho}\, \epsilon \, H_{\hat \mu \hat \nu \hat \rho} + (\textrm{fermion})^2 
    \eea

The Lagrangian for ${\cal D}=4$ $N=1$  Supergravity coupled with
one chiral supermultiplet, with no superpotential and the cannonical kinetic term
for the scalars is  
\cite{Cremmer:1978hn}
\begin{eqnarray} \nonumber
    L&=& - \frac{1}{2} R - \frac{1}{2} \bar { \psi}_{\hat \mu} \gamma^{\hat \mu \hat \nu \hat \rho} \left ( {\cal D}_{\hat \nu}
    +\frac{1}{8} ((\partial_{\hat \nu} z) z^\star- (\partial_{\hat \nu} z^\star) z) \gamma_5
   \right ){ \psi}_{\hat \rho} - \\ \nonumber
     &-&\frac{1}{2} (\partial_\mu z) (\partial_\mu z^\star)
    - \frac{1}{2} \bar \Xi \left ( \gamma^{\hat \mu} {\cal D}_{\hat \mu} -\frac{1}{8} ((\partial_{\hat \nu} z) z^\star- (\partial_{\hat \nu} z^\star) z)
    \right ) \Xi\\ \nonumber
    &+&(\textrm{fermion})^4
    \end{eqnarray}
    which is invariant under supersymmetry transformations
   \bea
    && \delta e_{\hat \mu}^\mu = \frac{1}{2}{\bar \epsilon} \, \gamma^{  \mu} \, \psi_{\hat \mu} \\ \nonumber
    && \delta z = \frac{1}{2} \, {\bar \epsilon} \, \Xi \\ \nonumber
    && \delta \psi_{\hat \mu} = \left ({\cal D}_{\hat \mu} +
    \frac{1}{8} ((\partial_{\hat \nu} z) z^\star- (\partial_{\hat \nu} z^\star) z) \right )  \epsilon + (\textrm{fermion})^2\\ \nonumber
    && \delta \Xi=  \frac{1}{2}(1+ \gamma_5)\, \gamma^{  \hat \mu} (\partial_{  \hat \mu} z) \epsilon + (\textrm{fermion})^2
    \eea 
with $\gamma_5= - \gamma_0\gamma_1\gamma_2\gamma_3$.

The Lagrangian for ${\cal D}=6$ $N=1$  Supergravity
coupled to one  $(1,0)$ tensor multiplet is
  \begin{eqnarray} \nonumber
    L&=& - \frac{1}{2} R - \frac{i}{2} \bar { \psi}_{\hat \mu} \gamma^{\hat \mu \hat \nu \hat \rho} {\cal D}_{\hat \nu} 
    { \psi}_{\hat \rho} 
    + \frac{1}{12} e^{2 \sqrt 2 \phi}\,H^{\hat \mu \hat \nu \hat \rho} H_{\hat \mu \hat \nu \hat \rho}
    +\\ \nonumber
    &+& \frac{i}{2} \bar \Xi \gamma^{\hat \mu} {\cal D}_{\hat \mu} \Xi 
    + \frac{1}{2}
     {\partial^{\hat \mu} \phi \, \partial_{ \hat \mu} \phi}
    - \frac{1 }{ \sqrt{2}}\bar { \Xi} \gamma^{ \hat \nu} \gamma^{ \hat \mu} \psi_{\hat \nu} \, \partial_{ \hat \nu} \phi  + \\ \nonumber
    &-& \frac{i}{24} 
    e^{\sqrt 2 \phi }H_{\hat \mu  \hat \nu \hat \rho} 
    (-\bar { \psi}^{ \hat \lambda} \gamma_{[\hat \lambda}
    \gamma^{ \hat \mu \hat \nu \hat \rho }
    \gamma_{\hat \tau]}
    { \psi}^{\hat \tau}
     +2i \bar { \psi}_{ \hat \lambda} \gamma^{ \hat \mu \hat \nu \hat \rho } \gamma^{\hat \lambda} \Xi
     - \bar \Xi \gamma^{ \hat \mu \hat \nu \hat \rho } \Xi)
     + \\ \nonumber
    &+&(\textrm{fermion})^4
    \end{eqnarray}
    where $H_{\hat \mu \hat \nu \hat \rho}= \partial_{[\hat \mu}B_{\hat \nu \hat \rho]}$ and ${\cal D}_{\hat \mu}$ is a covariant derivative
\be
{\cal D}_{{\hat \mu}} 
\Psi_{{\hat \nu}}^a =
\partial_{{\hat \mu}} \Psi_{{\hat \nu}}^a + 
\frac{1}{4} \omega_{{\hat \mu}}{}^{\rho \sigma} (\gamma_{\rho \sigma}
\Psi_{{\hat \nu}})^a
\ee
    The supersymmetry transformations with a local parameter $\epsilon^a(x)$ are
    \bea
    && \delta e_{\hat \mu}^\mu = -i
    {\bar \epsilon} \, \gamma^{  \mu} \, \psi_{\hat \mu} \\ \nonumber
    && \delta \phi = \frac{1}{\sqrt 2}  \, {\bar \epsilon} \, \Xi \\ \nonumber
    && \delta B_{\hat \mu \hat \nu} = 
    - \frac{i}{ 2}e^{- \sqrt 2 \phi}
    \,({\bar \epsilon} \, \gamma_{\hat \mu} \, \psi_{\hat \nu}
    - {\bar \epsilon} \, \gamma_{\hat \nu} \, \psi_{\hat \mu} 
    -i \, {\bar \epsilon} \, \gamma_{\hat \mu \hat \nu} \Xi) \\ \nonumber
    && \delta \psi_{\hat \mu} =\left ( {\cal D}_{\hat \mu} - \frac{1}{24} e^{\sqrt 2 \phi} 
    \gamma^{\hat \rho \hat \sigma  \hat \tau} \gamma_{\hat \mu}
    \, H_{\hat \rho \hat \sigma \hat \tau} 
    \right )  \epsilon
    + (\textrm{fermion})^2 \\
    && \delta \Xi=-
    \frac{i}{\sqrt 2}
 \gamma^{\hat \mu} \epsilon \, \partial_{\hat \mu} \, \phi
 - \frac{i}{1 2} e^{\sqrt 2 \phi} \gamma^{\hat \mu \hat \nu \hat \rho} \epsilon H_{\hat \mu \hat \nu \hat \rho}
+(\textrm{fermion})^2 
    \eea

We linearize around a flat background
   \be
e_{{\hat \mu}}^\mu(x) = \delta_{\hat \mu}^\mu + \frac{1}{2} h_{{\hat \mu}}^\mu(x),
\quad
e^{{\hat \mu}}_\mu(x) = \delta^{\hat \mu}_\mu - \frac{1}{2} h^{{\hat \mu}}_\mu(x),
\ee
    and consider a cubic Lagrangian with
    global   supersymmetry.

\providecommand{\href}[2]{#2}\begingroup\raggedright\endgroup

\end{document}